\documentclass[11pt]{article}
\usepackage{amsmath,lscape,amsfonts,amssymb,amsthm,verbatim,color,lscape}
\usepackage[figuresright]{rotating}
\usepackage{url}
\usepackage{hyperref}
\hypersetup{colorlinks,
citecolor=black,
filecolor=black,
linkcolor=black,
urlcolor=black,
pdftex}

\usepackage{times}
\usepackage{blindtext}

\usepackage{sectsty}
\subsectionfont{\normalsize\bf}
\sectionfont{\large\bf}

\usepackage[font={footnotesize,it}]{caption}
\def\p{\partial}

\def\Cbar{{\overline C}}

\def\x{\mathbf{x}}

\def\s{\sigma}

\def\X{\mathbf{X}}

\def\a{\alpha}
\def\g{\gamma}
\def\b{\beta}

\def\th{\theta}
\def\thbf{\boldsymbol{\theta}}

\def\Sbf{\boldsymbol{\Sigma}}

\def\bmu{\boldsymbol{\mu}}

\long\def\symbolfootnote[#1]#2{\begingroup
\def\thefootnote{\fnsymbol{footnote}}\footnote[#1]{#2}\endgroup}

\begin{document}
\title{A mixed effect model
for bivariate meta-analysis of diagnostic test accuracy studies using  a copula representation of the random effects distribution}

\date{}
\author{
Aristidis K. Nikoloulopoulos\footnote{{\small\texttt{A.Nikoloulopoulos@uea.ac.uk}}, School of Computing Sciences, University of East Anglia,
Norwich NR4 7TJ, UK}
}
\maketitle

\vspace{2ex}

\begin{abstract}
\baselineskip=13pt
\noindent Diagnostic test accuracy studies typically report the number of true positives, false positives,
true negatives and false negatives. There usually exists a negative association between the number of true
positives and true negatives, because studies that adopt less stringent
criterion for declaring a test positive invoke higher sensitivities and lower specificities.
A generalized linear mixed model (GLMM) is currently recommended to synthesize
diagnostic test accuracy studies.
We propose a copula mixed model for bivariate meta-analysis of diagnostic test accuracy studies.
Our general model includes the GLMM as a special case and can also operate on the original scale of sensitivity and specificity.
Summary receiver operating characteristic curves are deduced for the proposed model through quantile regression techniques and different characterizations of the  bivariate random effects distribution.
Our general
methodology is demonstrated with an extensive simulation study and illustrated by re-analysing the data of two  published meta-analyses.
Our study suggests that
there can be an improvement on GLMM in fit to data and makes the argument for moving to copula random effects models. Our modelling framework is implemented in the package {\tt CopulaREMADA} within the open source statistical environment {\tt R}.
\\\\
\noindent {\it Keywords:} {copula models; diagnostic tests; multivariate meta-analysis;  random effects models;  SROC, sensitivity/specificity.}
\end{abstract}

\baselineskip=12pt

\section{Introduction}
Synthesis of diagnostic test accuracy studies is  the most common medical application of multivariate meta-analysis \cite{JacksonRileyWhite2011,
MavridisSalanti13}.
Meta-analysis is broadly defined as the quantitative review of the results of related but independent studies      \cite{normand99}.
The purpose of a meta-analysis of diagnostic test accuracy studies is to combine information over different studies, and provide an integrated analysis that will have more statistical power to detect an accurate diagnostic test than an analysis based on a single study. Accurate diagnosis plays an important role in the disease control and prevention \cite{Ma-etal-2013}.

Diagnostic test accuracy studies observe the result of a gold standard procedure which defines the presence or absence of a decease and the result of  a diagnostic test.  They
typically report the number of true positives (diseased
people correctly diagnosed), false positives (non-diseased people incorrectly diagnosed as diseased),
true negatives and false negatives. As the
sensitivity (proportion of those with the disease) and specificity (proportion of those without the disease)
are estimated from different samples in each study (diseased and non-diseased
patients), they can be assumed to be independent so that the within-study correlations  are
set to zero \cite{MavridisSalanti13}. However, there may be a  negative between-studies association  which should be
accounted for. A negative
association between these quantities across studies is likely  because studies that adopt less stringent
criterion for declaring a test positive invoke higher sensitivities and lower specificities \cite{JacksonRileyWhite2011}.

In situations where studies compare a diagnostic test with its gold standard,
heterogeneity arises between studies due to the differences in disease prevalence,
study design as well as laboratory and other characteristics \cite{Chu-etal-2012}. Because of this heterogeneity,
a generalized linear mixed model (GLMM)
has been recommended in the biostatistics literature \cite{Chu&Cole2006,Arends-etal-2008,hamza-etal-2009,Ma-etal-2013} to synthesize information. Note in passing that it is  equivalent with  the hierarchical summary receiver operating characteristic model in Rutter and Gatsonis \cite{RutterGatsonis2001} for the case without covariates \cite{Harbord-etal-2007,Chu&Guo2009}.
The GLMM assumes independent binomial distributions for the true positives and true negatives, conditional on the latent pair of transformed (via a link function) sensitivity
and specificity  in each study. The  random effects (latent pair of transformed sensitivity
and specificity) are jointly analysed with a bivariate normal (BVN) distribution.

Chu {\it et al.}  \cite{Chu-etal-2012} propose an alternative  mixed model which operates on the original scale of sensitivity and specificity. The random effects follow the bivariate Sarmanov's \cite{Sarmanov1966} family of distributions with beta margins \cite{Lee-1996}. However, this random effects distribution has a limited range
of dependence and is inappropriate for general modelling unless the responses are
weakly dependent. Hence, this model is too restrictive in the context of diagnostic accuracy studies where strong (negative) dependence is likely.

We propose a copula mixed model as an extension of the GLMM and mixed model in Chu {\it et al.}  \cite{Chu-etal-2012} by rather using a copula representation of the random effects distribution  with normal and  beta margins, respectively.
Copulas are a useful way to model multivariate data as they account for the dependence structure and provide a flexible representation of the multivariate distribution. The theory and application of copulas have become important in finance, insurance and other areas, in order to deal with dependence in the joint tails. Here, we indicate that this can also be important in meta-analysis of diagnostic test accuracy studies.
Diagnostic test accuracy studies is a prime area of application for copula models, as the traditional assumption of multivariate normality is invalid in this context.

A copula approach for meta-analysis of diagnostic accuracy studies was recently proposed by Kuss {\it et al.}  \cite{kuss-etal-2013} who explored the use of  a copula model for observed discrete variables (number of true positives and true negatives) which have beta-binomial margins. This model is actually an approximation of a copula mixed model with beta margins for the latent pair of sensitivity and specificity.
Although, this approximation  can only be used under the unrealistic case that the number of observations in the respective study group of healthy and diseased probands  is the same for each study. In real data applications,
the number of true positives and negatives  do not have a common support over different studies, hence, one cannot conclude that there is a copula. The natural replicability is in the random effects probability for sensitivity and specificity.

The remainder of the paper proceeds as follows. Section \ref{stand-model-sec} summarizes the standard GLMM for synthesis of diagnostic test accuracy studies.
Section \ref{copula-mixed-model-sec} has a brief overview of
relevant copula theory and then introduces the copula mixed model for diagnostic test accuracy studies and discusses its relationship with existing mixed models.
 Section \ref{sec-families} discusses  suitable  parametric families of copulas for the copula mixed model, deduces summary receiver operating characteristic curves for the proposed model through quantile regression techniques and different characterizations of the  bivariate random effects distribution, and demonstrates that they can show the effect of different model assumptions.   Section \ref{miss-section} contains  small-sample  efficiency calculations
to  investigate the effect of misspecifying the random effects distribution on parameter estimators and standard errors and compare the proposed methodology to existing methods.
Section \ref{vuong-sec} summarizes the assessment of the proposed models using the Vuong's statistic \cite{vuong1989}, which is based on  sample difference in Kullback-Leibler divergence between two models and can be used to differentiate two  parametric models which could be non-nested.
Section \ref{sec-appl} presents applications of our methodology to four data frames with diagnostic accuracy data from binary test
outcomes. We conclude with some discussion in Section \ref{sec-discussion}, followed by a section with the software details and a technical Appendix.

\section{\label{stand-model-sec} The standard GLMM }

We first introduce the notation used in this paper. The focus is on two-level (within-study and between-studies) cluster data. The data are are $(y_{ij}, n_{ij}),\, i = 1, . . . ,N,\, j=1,2$, where $j$ is an index for the within study measurements and $i$ is an
index for the individual studies.
The data, for study $i$, can be summarized in a $2\times 2$ table with the number of true positives
($y_{i1}$), true negatives ($y_{i2}$), false negatives ($n_{i1}-y_{i1}$), and false positives ($n_{i2}-y_{i2}$); see Table \ref{2times2}.

\begin{table}[!h]
\caption{\label{2times2}Data from an individual study in a  $2\times 2$  table.}
\centering
\begin{footnotesize}
\begin{tabular}{ccc}
\hline
Test&\multicolumn{2}{c}{Disease (by gold standard)}\\
  & Yes & No \\
\hline
Positive & $y_{i1}$ & $n_{i2}-y_{i2}$ \\
Negative & $n_{i1}-y_{i1}$ & $y_{i2}$ \\\hline
Total & $n_{i1}$ & $n_{i2}$\\
\hline
\end{tabular}
\end{footnotesize}
\end{table}

The standard two-level model of meta-analysing diagnostic test accuracy studies     \cite{Chu&Cole2006,Harbord-etal-2007,Arends-etal-2008,hamza-etal-2009,Ma-etal-2013}  lies in the framework of mixed models \cite{Demidenko04}.
The within-study model assumes that the number of true positives $Y_{i1}$ and true negatives $Y_{i2}$ are conditionally independent and binomially distributed given $\X=\x$, where $\X=(X_1,X_2)$ denotes the  bivariate latent (random) pair of transformed sensitivity and specificity.  That is
\begin{eqnarray}\label{withinBinom}
Y_{i1}|X_{1}=x_1&\sim& \mbox{Binomial}\Bigl(n_{i1},l^{-1}(x_1)\Bigr);\nonumber\\
Y_{i2}|X_{2}=x_2&\sim& \mbox{Binomial}\Bigl(n_{i2},l^{-1}(x_2)\Bigr),
\end{eqnarray}
where $l(\cdot)$ is a link function such as the commonly used logit.
The between studies model assumes
that $\X$ is BVN distributed with mean vector $\bmu=\bigl(l(\pi_1),l(\pi_2)\bigr)^\top$ and variance covariance matrix
$\Sbf=\begin{pmatrix}
\sigma_1^2 &\rho\sigma_1\s_2\\
\rho\sigma_1\sigma_2 & \sigma_2^2
\end{pmatrix}$. That is
\begin{equation}\label{between}
\X
\sim
\mbox{BVN}
\bigl(\bmu,\Sbf\bigr).
\end{equation}
The models in (\ref{withinBinom}) and (\ref{between}) together specify a GLMM with joint likelihood
$$
L(\pi_1,\pi_2,\sigma_1,\sigma_2,\rho)=\prod_{i=1}^N\int\int
\prod_{j=1}^2g\Bigl(y_{ij};n_{ij},l^{-1}(x_j)\Bigr)\phi_{12}(x_1,x_2;\bmu,\Sbf)dx_1dx_2,
$$
where
$$g\bigl(y;n,\pi\bigr)=\binom{n}{y}\pi^y(1-\pi)^{n-y},\quad y=0,1,\ldots,n,\quad 0<\pi<1,$$
 is the binomial probability mass function (pmf) and $\phi_{12}(\cdot;\bmu,\Sbf)$ is the BVN density with mean vector $\bmu$  and variance covariance matrix $\Sbf$.
The parameters $\pi_1$ and $\pi_2$ are those of actual interest denoting the meta-analytic parameters for the sensitivity  and specificity, respectively, while the univariate parameters $\s_1^2$ and $\s^2_2$  are of secondary interest denoting the variability between studies.

\section{\label{copula-mixed-model-sec}The copula mixed model for diagnostic test accuracy studies  }
In this section, we  introduce the copula mixed model for diagnostic test accuracy studies and discuss its relationship with existing mixed models. Before that, the first subsection has some background on copula models.
In Subsection \ref{normal-parametrization} and Subsection \ref{beta-parametrization} a copula representation of the random effects distribution with normal and beta margins respectively is presented. We complete this section with details on maximum likelihood estimation.

\subsection{\label{overview}Overview and relevant background for copulas}
A copula is a multivariate cumulative distribution function (cdf) with uniform $U(0,1)$ margins \cite{joe97,nelsen06,joe2014}.
If $F_{12}$ is a bivariate cdf with univariate margins $F_1,F_2$,
then Sklar's \cite{sklar1959} theorem implies that there is a copula $C$ such that
\begin{equation}\label{copulacdf}
F_{12}(x_1,x_2)= C\Bigl(F_1(x_1),F_2(x_2)\Bigr).
\end{equation}
The copula is unique if $F_1,F_2$ are continuous, but not
if some of the $F_j$ have discrete components.
If $F_{12}$ is continuous and $(X_1,X_2)\sim F_{12}$, then the unique copula
is the distribution of $(U_1,U_2)=\left(F_1(X_1),F_2(X_2)\right)$ leading to
  $$C(u_1,u_2)=F_{12}\Bigl(F_1^{-1}(u_1),F_2^{-1}(u_2)\Bigr),
  \quad 0\le u_j\le 1, j=1,2,$$
where $F_j^{-1}$ are inverse cdfs. In particular, if $\Phi_{12}(\cdot;\rho)$
is the BVN cdf with correlation $\rho$ and
standard normal margins, and $\Phi$ is the univariate standard normal cdf,
then the BVN copula is
$$
C(u_1,u_2)=\Phi_{12}\Bigl(\Phi^{-1}(u_1),\Phi^{-1}(u_2);\rho\Bigr).
$$
The power of copulas for dependence modelling is due to
the dependence structure being considered separate from the univariate
margins; see e.g., \cite[Section 1.6]{joe97}.
If $C(\cdot;\theta)$ is a parametric
family of copulas and $F_j(\cdot;\eta_j)$ is a parametric model for the
$j$th univariate margin, then
  $$C\Bigl(F_1(x_1;\eta_1),F_2(x_2;\eta_2);\theta\Bigr)$$
is a bivariate parametric model with univariate margins $F_1,F_2$.
For copula models, the variables can be continuous or discrete   \cite{nikoloulopoulos&joe12}.

\subsection{\label{normal-parametrization}The copula mixed  model for the latent pair of transformed sensitivity and specificity}
Here we generalize the GLMM by  proposing a model that links the two random effects using a copula function rather than the BVN distribution.

The within-study model is the same as in the standard GLMM; see (\ref{withinBinom}).
The stochastic representation of the between studies model takes the form
\begin{equation}\label{copula-between-norm}
\Bigl(\Phi\bigl(X_1;l(\pi_1),\s_1^2\bigr),\Phi\bigl(X_2;l(\pi_2),\s_2^2\bigr)\Bigr)\sim C(\cdot;\th),
\end{equation}
where $C(\cdot;\th)$ is a parametric family of copulas with dependence parameter $\th$ and $\Phi(\cdot;\mu,\s^2)$ is the cdf of the  N($\mu,\s^2$) distribution.
The joint density $f_{12}(x_1,x_2)$ of the transformed latent proportions can be derived as a double partial derivative of the cdf in (\ref{copulacdf})
\begin{eqnarray}\label{jointdensityNCMM}
f_{12}(x_1,x_2;\pi_1,\pi_2,\s_1,\s_2,\th)=\frac{\p C\Bigr(\Phi\bigr(x_1;l(\pi_1),\s_1^2\bigl),\Phi\bigr(x_2;l(\pi_2),\s_2^2\bigl);\th\Bigl)}{\p x_1\p x_2}&&\\
=c\Bigr(\Phi\bigl(x_1;l(\pi_1),\s_1^2\bigr),\Phi\bigl(x_2;l(\pi_2),\s_2^2\bigr);\th\Bigl)\phi\bigl(x_1;l(\pi_1),\s_1^2\bigr)\phi\bigl(x_2;l(\pi_2),\s_2^2\bigr),&&\nonumber
\end{eqnarray}
where $c(u_1,u_2;\th)=\p^2 C(u_1,u_2;\th)/\p u_1\p u_2$  and $\phi(\cdot;\mu,\s^2)$ is the copula and N($\mu,\s^2$)  density, respectively.
The models in (\ref{withinBinom}) and (\ref{copula-between-norm}) together specify a copula mixed  model with joint likelihood

\begin{eqnarray}
\label{mixed-cop-likelihood}
L(\pi_1,\pi_2,\s_1,\s_2,\th)&=&\prod_{i=1}^N\int_{-\infty}^{\infty}\int_{-\infty}^{\infty}
\prod_{j=1}^2g\Bigl(y_{ij};n_{ij},l^{-1}(x_j)\Bigr)c\Bigl(\Phi\bigl(x_1;l(\pi_1),\s_1^2\bigr),\\
&&\qquad\qquad\Phi\bigl(x_2;l(\pi_2),\s_2^2\bigr);\th\Bigr)\prod_{j=1}^2\phi\bigl(x_j;l(\pi_j),\s_j^2\bigr)dx_1dx_2\nonumber\\
&=&\prod_{i=1}^N\int_{0}^{1}\int_{0}^{1}
\prod_{j=1}^2g\Bigl(y_{ij};n_{ij},l^{-1}\bigl(\Phi^{-1}(u_j;l(\pi_j),\s_j^2)\bigr)\Bigr)c(u_1,u_2;\th)du_1du_2.\nonumber
\end{eqnarray}

It is important to note that the copula parameter $\th$ is a parameter of the random effects model and it is separated from the univariate parameters. The univariate parameters $\pi_1$ and $\pi_2$ are those of actual interest denoting the meta-analytic parameters for the sensitivity and specificity, while the univariate parameters $\sigma_1^2$ and $\sigma_2^2$  are of secondary interest expressing the variability between studies.

\subsubsection{Relationship with the GLMM}

In this subsection, we show what happens when  the bivariate copula is the BVN copula.
The resulting model is the same as the GLMM.

The BVN copula density is
$$c(u_1,u_2;\rho)=\frac{1}{\sqrt{1-\rho^2}}\exp\left(\frac{z_1^2+z^2_2-2\rho z_1z_2}{2\sqrt{1-\rho^2}}\right)\exp\left(\frac{z_1^2+z_2^2}{2}\right),$$
where $z_j=\Phi^{-1}(u_j),\,j=1,2$. Then for $u_j=\Phi\bigl(x_j;l(\pi_j),\s_j^2\bigr)$ we have  $z_j=\bigl(x_j-l(\pi_j)\bigr)/\s_j,\,j=1,2$. Hence, the joint density in (\ref{jointdensityNCMM}) becomes
\begin{eqnarray*}
f_{12}(x_1,x_2;\pi_1,\pi_2,\s_1,\s_2,\rho)&=&\frac{1}{2\pi\s_1\s_2\sqrt{1-\rho^2}}\exp\Bigl[\frac{1}{2\sqrt{1-\rho^2}}
\Bigl\{\frac{\bigl(x_1-l(\pi_1)\bigr)^2}{2\s_1^2}+\\
&&\frac{\bigl(x_2-l(\pi_2)\bigr)^2}{2\s_2^2}-2\rho \frac{\bigl(x_1-l(\pi_1)\bigr)\bigl(x_2-l(\pi_2)\bigr)}{\s_1\s_2}\Bigr\}\Bigr],
\end{eqnarray*}
which apparently is the BVN density $\phi_{12}(x_1,x_2;\bmu,\Sbf)$.

\subsection{\label{beta-parametrization}The copula mixed  model for the latent pair of sensitivity and specificity}

The within-study model also assumes that the number of true positives $Y_{i1}$ and true negatives $Y_{i2}$ are conditionally independent and binomially distributed given $\X=\x$, where $\X=(X_1,X_2)$ denotes the  bivariate latent random pair of sensitivity and specificity.
That is
\begin{eqnarray}\label{withinBinom2}
Y_{i1}|X_{1}=x_1&\sim& \mbox{Binomial}(n_{i1},x_1);\nonumber\\
Y_{i2}|X_{2}=x_2&\sim& \mbox{Binomial}(n_{i2},x_2).
\end{eqnarray}
So one  does not have to transform the latent sensitivity and specificity and can work on the original scale.
The Beta($\a,\b$) distribution can be used for the marginal modeling of
the latent proportions and its density is
$$f(x;\a,\b)=\frac{x^{\a-1}(1-x)^{\b-1}}{B(\a,\b)},\quad 0<x<1, \quad \a,\b>0.$$
In the sequel we will use the
Beta($\pi,\gamma$) parametrization, where $\pi=\frac{\a}{\a+\b}$ (mean parameter) and $\g=\frac{1}{\a+\b+1}$ (dispersion parameter).

The stochastic representation of the between studies model is
\begin{equation}\label{copula-between}
\Bigl(F(X_1;\pi_1,\g_1),F(X_2;\pi_2,\g_2)\Bigr)\sim C(\cdot;\th),
\end{equation}
where $C(\cdot;\th)$ is a parametric family of copulas with dependence parameter $\th$ and $F(\cdot;\pi,\g)$ is the cdf of the the Beta($\pi,\g$) distribution.
The models in (\ref{withinBinom2}) and (\ref{copula-between}) together specify a copula mixed  model with joint likelihood

\begin{eqnarray}\label{beta-mixed-cop-likelihood}
L(\pi_1,\pi_2,\g_1,\g_2,\th)&=&\prod_{i=1}^N\int_0^1\int_0^1
\prod_{j=1}^2g(y_{ij};n_{ij},x_j)c\Bigl(F(x_1;\pi_1,\g_1),F(x_2;\pi_2,\g_2);\th\Bigr)\nonumber\\
&&\qquad\qquad\qquad\times\prod_{j=1}^2f(x_j;\pi_j,\g_j)dx_1dx_2\\
&=&\prod_{i=1}^N\int_0^1\int_0^1
\prod_{j=1}^2g\bigl(y_{ij};n_{ij},F^{-1}(u_j;\pi_j,\g_j)\bigr)c(u_1,u_2;\th)du_1du_2.\nonumber
\end{eqnarray}

As before, the copula parameter $\th$ is a parameter of the random effects model and it is separated from the univariate parameters, the univariate parameters $\pi_1$ and $\pi_2$ are the meta-analytic parameters for the sensitivity and specificity, but, now  $\gamma_1$ and $\gamma_2$  express the variability between studies.

\subsubsection{Relationship with existing models}

Chu {\it et al.} \cite{Chu-etal-2012}, instead of using a copula for the random effects distribution or a copula density for
$\X$ in (\ref{beta-mixed-cop-likelihood}), use the Sarmanov's \cite{Sarmanov1966} family of bivariate densities
$$
f_{12}(x_1,x_2)=f_1(x_1)f_2(x_2)\Bigl(1+\th\psi_1(x_1)\psi_2(x_2)\Bigr),
$$
where $f_j(\cdot)$ is the marginal density of $X_j$, $\psi_j(\cdot)$ is a bounded non-constant function such as
$\int_{-\infty}^{\infty} f_j(x)$ $\psi_j(x) dx = 0$,
and $1+\th\psi_1(x_1)\psi_2(x_2)\geq0$ for all $x_1,x_2$.
For the Sarmanov's densities if one uses $\psi_j=1-2F_j(x_j),\,j=1,2$, then the Farlie--Gumbel--Morgenstern  copula (density) is obtained.
However in \cite{Chu-etal-2012}, ``kernels'' of the type $\psi_j(x_j)=x_j-E(X_j)$ are considered as in \cite{Lee-1996}.
The advantage of this choice is
that the
corresponding likelihood function has a closed form, since the product of integrals can
be evaluated analytically. The joint likelihood takes the form

\begin{eqnarray*}\label{sarmanov-likelihood}
L(\pi_1,\pi_2,\g_1,\g_2,\th)&=&\prod_{i=1}^N\int_0^1\int_0^1
\prod_{j=1}^2g(y_{ij};n_{ij},x_j)f(x_j;\pi_j,\g_j)\Bigl(1+\th\prod_{j=1}^2\bigl(x_j-\pi_j\bigr)\Bigr)
dx_1dx_2\nonumber\\
&=&\prod_{i=1}^N\prod_{j=1}^2h(y_{ij};n_{ij},\pi_j,\g_j)\Bigl(1+\th\prod_{j=1}^2\frac{y_{ij}-n_{ij}\pi_j}{\g_j^{-1}+n_{ij}-1}\Bigr),
\end{eqnarray*}
where $$h(y;n,\pi,\g)=\binom{n}{y}\frac{B\Bigl(y+\pi/\g-\pi,n-y+(1 - \pi)(1 - \g)/\g\Bigr)}{B\Bigl(\pi/\g-\pi,(1 - \pi)(1 - \g)/\g\Bigr)},\, y=0,1,\ldots,n,\, 0<\pi,\g< 1,$$
is the pmf of a Beta-Binomial($n,\pi,\g$) distribution with mean $n\pi$ and variance $n\pi(1-\pi)\bigl(1+(n-1)\g\bigr)$.
The disadvantage of this mixed model is that
the   Sarmanov's  density with beta margins in \cite{Lee-1996} has a limited range
of dependence and is inappropriate for general modeling unless the responses are
weakly dependent.

Kuss {\it et al.} \cite{kuss-etal-2013}  proposed a copula model with beta-binomial margins in this context. This model is actually an approximation of the copula mixed model with beta margins for the latent pair of sensitivity and specificity in (\ref{withinBinom2}) and (\ref{copula-between}).
They attempt to approximate the likelihood  in (\ref{beta-mixed-cop-likelihood})  with the likelihood of a copula model for observed discrete variables which have beta-binomial margins.

The approximation that they suggest is
$$
L(\pi_1,\pi_2,\g_1,\g_2,\th)\approx
\prod_{i=1}^Nc\Bigl(H(y_{i1};n_{i1},\pi_1,\g_1),H(y_{i2};n_{i2},\pi_2,\g_2);\th\Bigr)\prod_{j=1}^2 h(y_{ij};n_{ij},\pi_j,\g_j),
$$
where $H(\cdot;n,\pi,\g)$ is the cdf of the the Beta-Binomial($n,\pi,\g$) distribution.
In their approximation the authors also treat the observed variables which have beta-binomial distributions as being continuous, and model them under the theory for copula models with continuous margins. Kuss {\it et al.} \cite{kuss-etal-2013}, referring to Genest and Ne\v{s}lehov\'{a}    \cite{genest&johanna07}, claim that there are problems on applying copula to discrete data especially in extreme cases with very small numbers of support points for the discrete marginal distributions. Genest and Ne\v{s}lehov\'{a}        \cite{genest&johanna07} only warn against estimation for discrete-margined copula models using rank-based methods, instead recommending maximum likelihood estimation. Essentially, Genest {\it et al.} \cite{genest&nikoloulopoulos&rivest08}  apply copula  models to multivariate binary data (the extreme case of discreteness) and call on composite likelihood techniques for estimation.
Multivariate copulas for discrete response data have been in use for a considerable length of time,  e.g., in Joe  \cite{joe97}, and earlier for some simple copula models. Several examples of copula models for multivariate discrete data can be found in the literature; see e.g.,  \cite{Nikoloulopoulos&karlis07BIN} for an application in biostatistics and \cite{Nikoloulopoulos2013a} for a survey of copula models and methods for multivariate discrete response data.

However, the main problem in   \cite{kuss-etal-2013} is that the approximation  (even if treating the observed variables which have beta-binomial distributions as being discrete)  can only be used under the unrealistic case that the number of observations in the respective study group of healthy and diseased probands $n_{ij}$ is the same for each study $i$. In real data applications,
the discrete  $Y_{ij}$ do not have a common support over different studies or $i$, hence, one cannot conclude that there is a copula for $(Y_{i1},Y_{i2})$ that applies when the $n_{ij}$ vary with different studies $i$. The natural replicability is in the random effects probability for sensitivity and specificity.

\subsection{\label{computation}Maximum likelihood estimation and computational details}

Estimation of the model parameters $(\pi_1,\pi_2,\s_1,\s_2,\th)$ and $(\pi_1,\pi_2,\g_1,\g_2,\th)$   can be approached by the standard maximum likelihood (ML) method, by maximizing the logarithm of the joint likelihood in (\ref{mixed-cop-likelihood})  and (\ref{beta-mixed-cop-likelihood}), respectively.
The estimated parameters can be obtained by
using a quasi-Newton \cite{nash90} method applied to the logarithm of the joint likelihood.
This numerical  method requires only the objective
function, i.e.,  the logarithm of the joint likelihood, while the gradients
are computed numerically and the Hessian matrix of the second
order derivatives is updated in each iteration. The standard errors (SE) of the ML estimates can be also obtained via the gradients and the Hessian computed numerically during the maximization process. Assuming that the usual regularity conditions \cite{serfling80} for
asymptotic maximum likelihood theory hold for the bivariate model
as well as for its margins we have that ML estimates are
asymptotically normal. Therefore one can build Wald tests to
statistically judge any effect.

For mixed models of the form with joint likelihood as in (\ref{mixed-cop-likelihood})  and (\ref{beta-mixed-cop-likelihood}), numerical evaluation of the joint pmf is easily done with the following steps:

\begin{enumerate}
\itemsep=10pt
\item Calculate Gauss-Legendre  quadrature points $\{u_q: q=1,\ldots,n_q\}$
and weights $\{w_q: q=1,\ldots,n_q\}$ in terms of standard uniform; see e.g., \cite{Stroud&Secrest1966}.
\item Convert from independent uniform random variables $\{u_{q_1}: q_1=1,\ldots,n_q\}$ and $\{u_{q_2}: q_2=1,\ldots,n_q\}$ to dependent uniform random variables $\{u_{q_1}: q_1=1,\ldots,n_q\}$ and $\{C^{-1}(u_{q_2}|u_{q_1};\th): q_1=q_2=1,\ldots,n_q\}$ that have distribution $C(\cdot;\th)$.
The inverse of the conditional distribution $C(v|u;\th)=\partial C(u,v;\th)/\partial u$ corresponding to the copula $C(\cdot;\th)$ is used  to achieve this.
\item Numerically evaluate the joint pmf, e.g.,
$$\int_0^1\int_0^1
\prod_{j=1}^2g\bigl(y_{j};n_{j},F^{-1}(u_j;\pi_j,\g_j)\bigr)c(u_1,u_2;\th)du_1du_2$$
in a double sum:
$$\sum_{q_1=1}^{n_q}\sum_{q_2=1}^{n_q}w_{q_1}w_{q_2}
g\bigl(y_{1};n,F^{-1}(u_{q_1};\pi_j,\g_j)\bigr)g\Bigl(y_{2};n,F^{-1}\bigl(C^{-1}(u_{q_2}|u_{q_1};\th);\pi_j,\g_j\bigr)\Bigr).$$
\end{enumerate}

With Gauss-Legendre quadrature, the same nodes and weights
are used for different functions;
this helps in yielding smooth numerical derivatives for numerical optimization via quasi-Newton \cite{nash90}.
Our comparisons show that $n_q=15$ is adequate with good precision to at least at four decimal places; hence it also provides the advantage of fast implementation.

To sum up,  our mixed  effect model for meta-analysis of diagnostic test accuracy studies using  a copula representation of the random effects distribution with a double integral
is straightforward computationally.
Note in passing that the linear mixed model in \cite{Reitsma-etal-2005} can also provide handy computations, but it has limitations due to the use of continuity correction and normal approximation \cite{Chu&Cole2006,Ma-etal-2013}.

\section{\label{sec-families}Choices of parametric families of copulas}
In our candidate set, families that have
different strengths of tail behaviour (see e.g., \cite{heffernan00}) are included. In the descriptions below, a bivariate copula $C$ is {\it reflection symmetric}
if its density
 satisfies $c(u_1,u_2)=c(1-u_1,1-u_2)$ for all $0\leq u_1,u_2\leq 1$.
Otherwise, it is reflection asymmetric often with more probability in the
joint upper tail or joint lower tail. {\it Upper tail dependence} means
that $c(1-u,1-u)=O(u^{-1})$ as $u\to 0$ and {\it lower tail dependence}
means that $c(u,u)=O(u^{-1})$ as $u\to 0$.
If $(U_1,U_2)\sim C$ for a bivariate copula $C$, then $(1-U_1,1-U_2)\sim
C_{180^\circ}$, where $C_{180^\circ}(u_1,u_2)=u_1+u_2-1+C(1-u_1,1-u_2)$ is the survival (or rotated by 180 degrees) copula of $C$; this ``reflection"
of each uniform $U(0,1)$ random variable about $1/2$ changes the direction
of tail asymmetry.
\begin{itemize}
\itemsep=10pt
\item
Reflection symmetric copulas with tail independence satisfying
$C(u,u)=O(u^2)$ and $\Cbar(1-u,1-u)=O(u^2)$ as $u\to 0$,
such as the Frank copula with inverse conditional cdf

$$C^{-1}(v|u;\th)=
-\frac{1}{\theta}\log\left[\frac{1-(1-e^{-\th})}{(v^{-1}-1)e^{-\th u}+1}\right]
,\quad \theta \in (-\infty,\infty)\setminus\{0\}.$$
\item
Reflection symmetric copulas with intermediate tail dependence
\cite{Hua-joe-11} such as the BVN copula, which satisfies
$C(u,u,\th)=O(u^{2/(1+ \th)}(-\log u)^{-\th/(1+\th)})$ as
$u\to 0$ with inverse conditional cdf
$$C^{-1}(v|u;\th)=\Phi\Bigl(\sqrt{1-\rho^2}\Phi^{-1}(v)+\rho\Phi^{-1}(u)\Bigr),\quad \theta \in [-1,1].$$
\item
Reflection asymmetric copulas with lower tail dependence only such as the Clayton copula with inverse conditional cdf
$$C^{-1}(v|u;\th)=
\Bigl\{(v^{-\theta/(1+\theta)}-1)u^{-\th}+1\Bigr\}^{-1/\theta}
,\quad \th\in (0,\infty).$$

\item
Reflection asymmetric copulas with  upper tail dependence only such as the
rotated by 180 degrees Clayton copula with
inverse conditional cdf
$$C^{-1}(v|u;\th)=
1-\Bigl[\bigl\{(1-v)^{-\theta/(1+\theta)}-1\bigr\}(1-u)^{-\th}+1\Bigr]^{-1/\theta}
,\quad \th\in (0,\infty).$$
\end{itemize}

The Frank and BVN
copulas interpolate from the Fr\'echet lower (perfect negative dependence) to the
Fr\'echet upper (perfect positive dependence) bound, and, thus they are sufficient from bivariate studies on diagnostic accuracy where negative dependence  between the number of true positives and true negatives is expected.
The Clayton copula belongs in the Archimedean class of copulas. Archimedean copulas, see e.g. \cite{joe97},
have the form,
\begin{equation} \label{permutation}
  C(u_1, u_2\, ;\, \theta)=\phi\left(\phi^{-1}(u_1\, ;\,\theta)+\phi^{-1}(u_2\, ;\,\theta)\, ;\, \theta\right),
\end{equation}
where the generator $\phi(u\, ;\,\theta)$ is the Laplace transform (LT) of a univariate
family of distributions of positive random variables indexed by the
parameter $\theta$, such that
$\phi(\cdot)$ and its inverse have closed forms.
The Clayton copula  interpolates from the independence ($\theta\to 0$) to the
Fr\'echet upper (comonotonic copula) bound ($\theta\to \infty$).
For extension of the Laplace transform  for $\theta\in [-1,0)$, the Clayton family extends to countermonotonicity ($\theta\to -1$). However this extension is not generally useful for applications because the support of (\ref{permutation}) is not all of $(0,1)^2$ \cite[page 109]{joe97}. Negative dependence in Clayton copulas can be introduced by applying decreasing
transformations to the ``oppositely'' ordered variables. If
$(U_1,U_2)\sim C$ where $C$ is a copula with positive
dependence, one could always get some negative dependence, by supposing $C_{90^\circ}$ is the copula of
$(U_1,1 -U_2)$ (rotation by 90 degrees) or $C_{270^\circ}$ the copula of  $(1-U_1,U_2)$ (rotation by 270 degrees).
So it is worthwhile to rotate the Clayton copula by 90 and 270 degrees to model negative dependence. These rotated copulas
interpolate from the Fr\'echet lower (perfect negative dependence) ($\theta\to \infty$) to the
independence  ($\theta\to0$).
{\it Negative upper-lower tail dependence} means
that $c(1-u,u)=O(u^{-1})$ as $u\to 0$ and {\it negative lower-upper tail dependence}
means that $c(u,1-u)=O(u^{-1})$ as $u\to 0$ \cite{joe2010b}. So in order to model negative (tail) dependence the choices are:
\begin{itemize}
\itemsep=10pt
\item Reflection asymmetric copula family with negative upper-lower tail dependence,
such as the rotated by 90 degrees Clayton copula with inverse conditional cdf
$$C^{-1}(v|u;\th)=
\Bigl\{(v^{\theta/(1-\theta)}-1)(1-u)^{\th}+1\Bigr\}^{1/\theta}
,\quad \th\in (0,\infty).$$

\item Reflection asymmetric copula family with negative lower-upper tail dependence,
such as the as the rotated by 270 degrees Clayton copula with inverse conditional cdf
$$C^{-1}(v|u;\th)=1-
\Bigl[\bigl\{(1-v)^{\theta/(1-\theta)}-1\bigr\}u^{\th}+1\Bigr]^{1/\theta}
,\quad \th\in (0,\infty).$$
\end{itemize}

For this paper, the above copula families are sufficient for the applications in Section \ref{sec-appl}, since tail dependence is a property to consider when choosing amongst different families of copulas and the concept of upper/lower tail dependence is one way to differentiate families. Nikoloulopoulos and Karlis \cite{Nikoloulopoulos&karlis08CSDA}
have shown that it is hard to choose a copula with similar properties from real data, since
copulas with similar (tail) dependence properties provide similar fit.
Kuss {\it et al.}  \cite{kuss-etal-2013} used, in addition to these copulas,  the Placket copula.
Plackett copula  is a reflection symmetric copula \cite[pages 221-22]{joe97} with tail independence \cite[page 215]{nelsen06} (not reflection asymmetric copula as stated in  \cite{kuss-etal-2013}) and is not used here since we have included another choice of copulas with similar properties i.e., the Frank copula.

\subsection{\label{sec-SROC}Summary receiver operating characteristic   curves}
Rutter and Gatsonis \cite{RutterGatsonis2001} proposed a hierarchical summary receiver operating characteristic (SROC) curve  which for some cases is the same with the corresponding GLMM  SROC curve \cite{Chu&Guo2009}.   For the GLMM model, the model parameters control the shape of the SROC curve. The GLMM SROC curve can be obtained through a characterization of the estimated bivariate normal distribution by a line        \cite{Chu&Guo2009,Chu-etal-2010,Chu-etal-2012}. Based on the bivariate normality of the random effects, the expected sensitivity for a chosen specificity in the transformed scale is given   in a closed form:
\begin{equation}
E[X_1|X_2=x_2]=[l(\pi_1)-\rho l(\pi_2)\s_1/\s_2]+\rho l(x_2)\s_1/\s_2.
\end{equation}
In general, however, $E[X_1|X_2=x_2]$ is not in closed form  and thus does not have simple expressions in terms of distribution functions and copulas.

An alternative to the mean for specifying ``typical" values of $X_1$ for each value of $X_2$ is the median, which leads to the notion of median regression  of $X_1$ on $X_2$.  For $x_2$ in range of $X_2$, let $x_1:=\widetilde{x}_1(x_2)$ denote a solution to the equation $\Pr(X_1\leq x_1|X_2=x_2)=1/2$. Then the scatter plot of $\widetilde{x}_1(x_2)$ and $x_2$ is the median regression curve of $X_1$ on $X_2$.

For  copula  models,
median regression curves \cite[pages 217--218]{nelsen06}   can be easily calculated, since

$$\Pr(X_1\leq x_1|X_2=x_2)=\Pr\bigl(U_1\leq F_1(x_1)|U_2=F_2(x_2)\bigr)=C(u_1|u_2)\Bigg\|_{\begin{array}{c}u_1=F_1(x_1)\\
u_2=F_2(x_2)
\end{array}},$$
but their shape  also depends on the choice of bivariate copulas. Furthermore, as emphasized in \cite{Rucker-schumacher-2009}, since there is no unique definition of  a SROC curve, it is preferable and will make more sense  to deduce confidence regions as well.
  To this end in addition of using just median regression curves we will also exploit the use of quantile regression curves with a focus on high ($q=0.99$) and low quantiles ($q=0.01$) which are strongly associated with the upper and lower tail dependence imposed from each parametric family of copulas. These can be also seen as confidence regions of the median regression SROC curve. Note that Kendall's tau only accounts for the dependence dominated by the middle of the data, and it is expected to be similar amongst  different families of copulas. However, the tail dependence varies, as explained in Section \ref{sec-families}, and is a property to differentiate amongst different families of copulas.

To find the quantile  regression curves:
\begin{enumerate}
\itemsep=10pt
\item Set $C(u_1|u_2;\th)=q$.
\item Solve for the quantile regression curve $u_1:=\widetilde{u}_1(u_2,q;\th)=C^{-1}(q|u_2;\th)$.
\item For $j=1,2$ replace $u_j$ by $F_j(x_j;\pi_j,\g_j)$ for beta margins or $\Phi_j\bigl(x_j;l(\pi_j),\s_j\bigr)$ for normal margins.
\item Plot $x_1:=\widetilde{x}_1(x_2,q)$ versus $x_2$.
\end{enumerate}

Of course, the quantile regression curve $x_2:=\widetilde{x}_2(x_1,q)$
of $X_2$ on $X_1$ is defined similarly.
However, there is no priori reason to regress $x_1$ on $x_2$ instead of the other way around \cite{Arends-etal-2008}. In fact, if one wants to reserve the nature of a bivariate response instead of a univariate response along with a covariate, then a contour graph can be easily plotted. The contour plot can be seen as the predictive region (analogously  to \cite{Reitsma-etal-2005}) of the estimated pair of sensitivity and specificity. However, the resulted shape of the prediction region is not depended   on the assumption of bivariate normality for the random effects.

\begin{figure}[!h]
\begin{center}
\begin{footnotesize}
\begin{tabular}{|cc|}
\hline

BVN& Frank\\\hline
\includegraphics[width=0.3\textwidth]{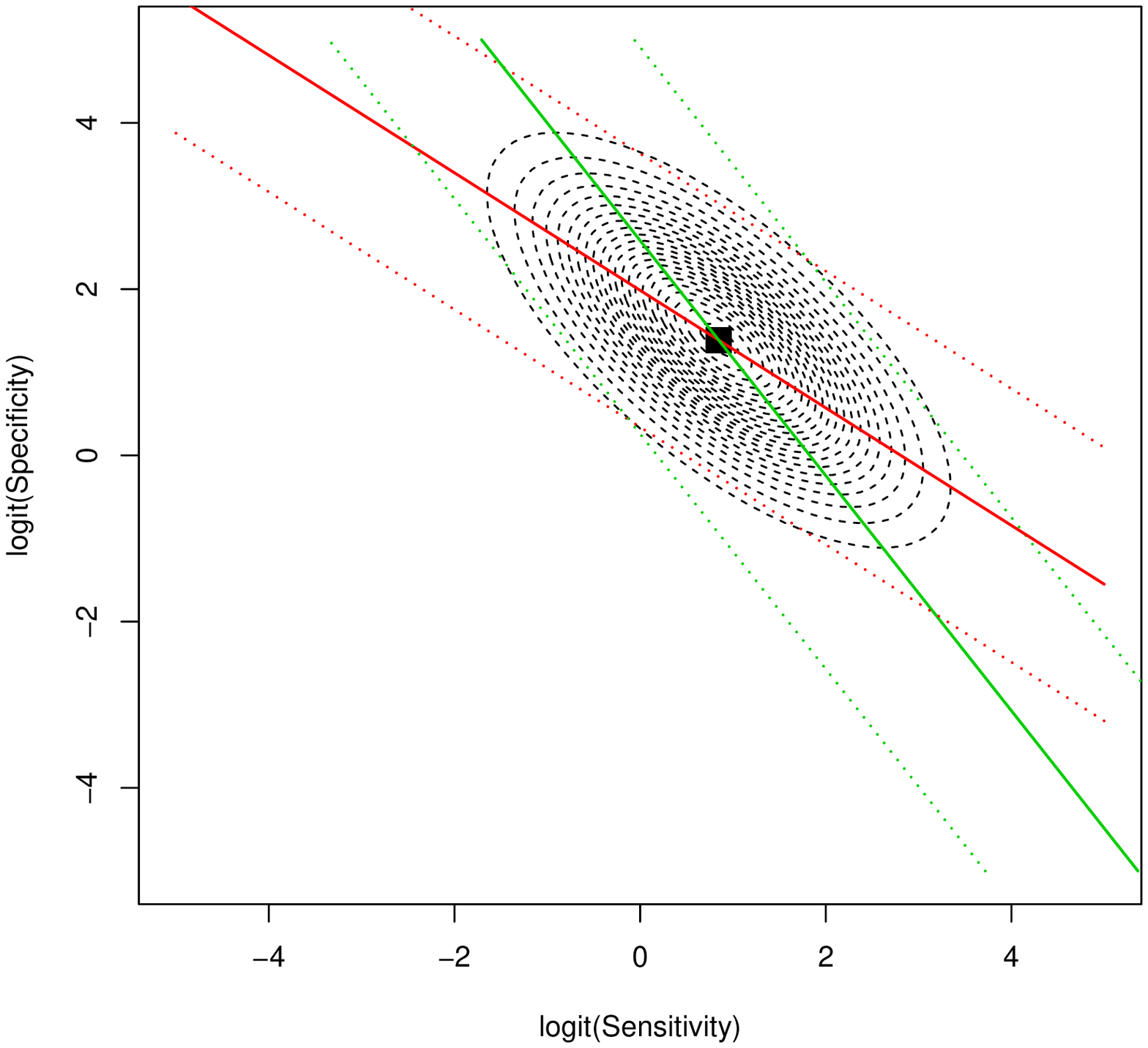}
&

\includegraphics[width=0.3\textwidth]{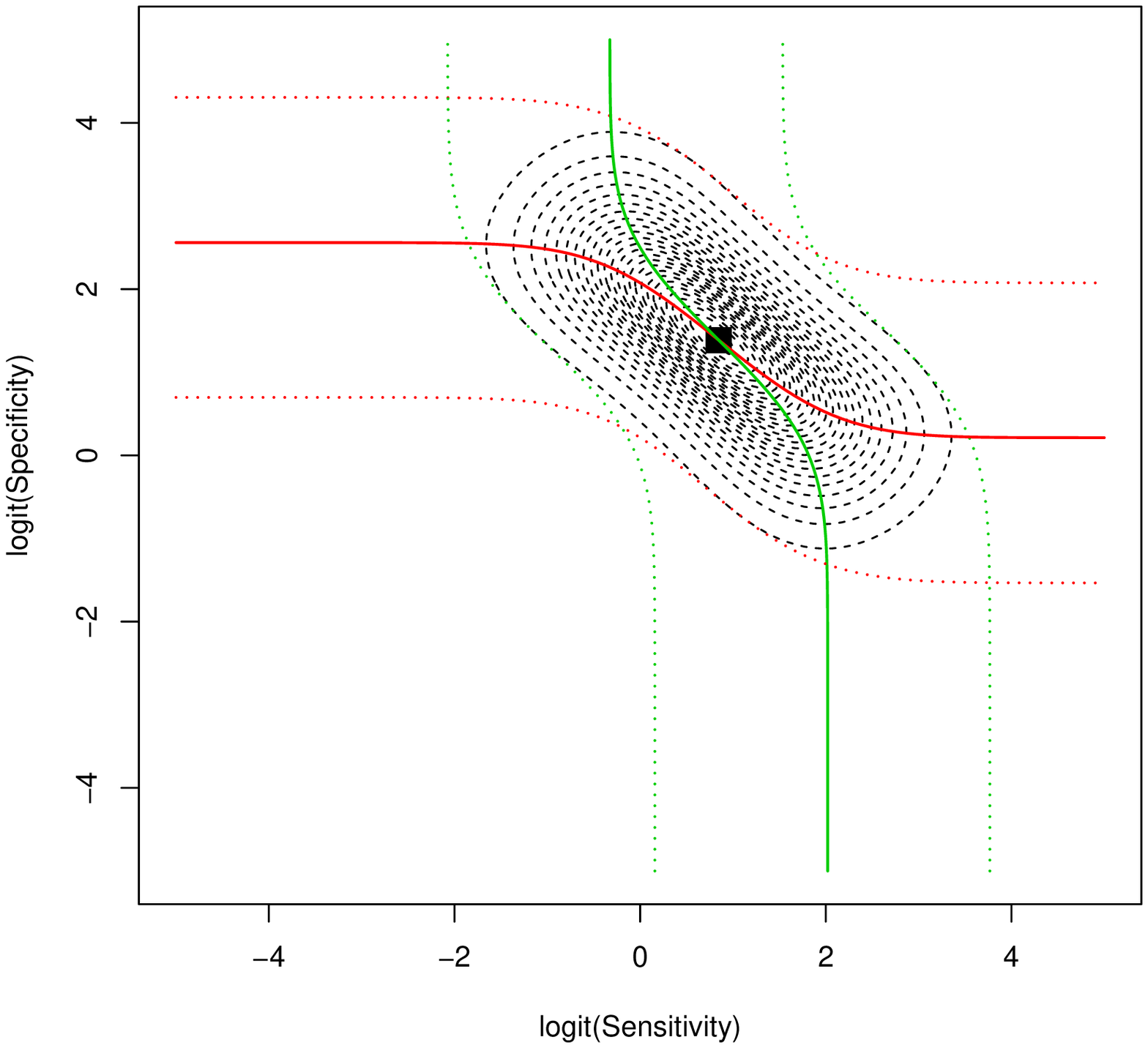}\\\hline
Clayton by 90& Clayton by 270\\\hline
\includegraphics[width=0.3\textwidth]{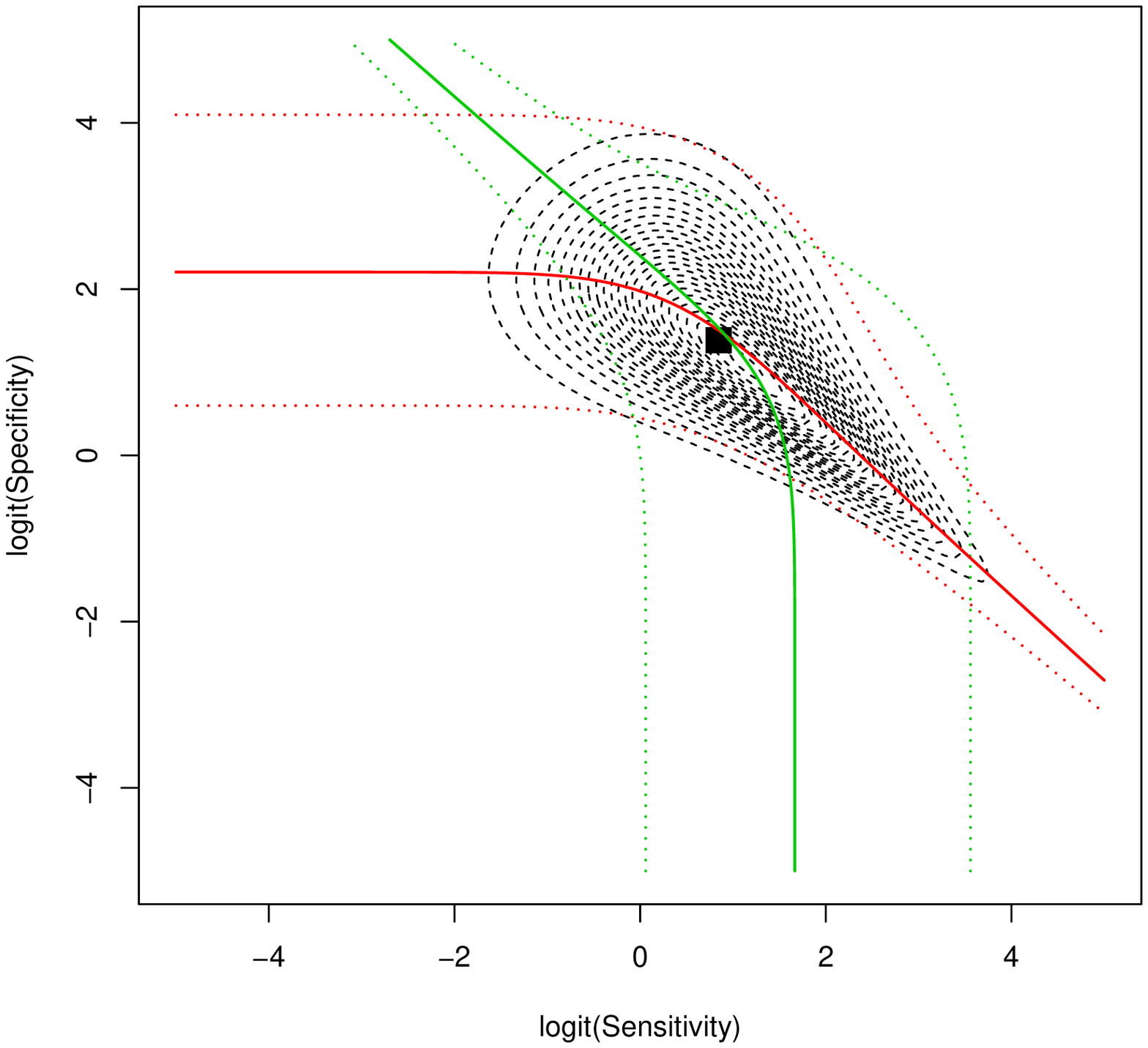}
&
\includegraphics[width=0.3\textwidth]{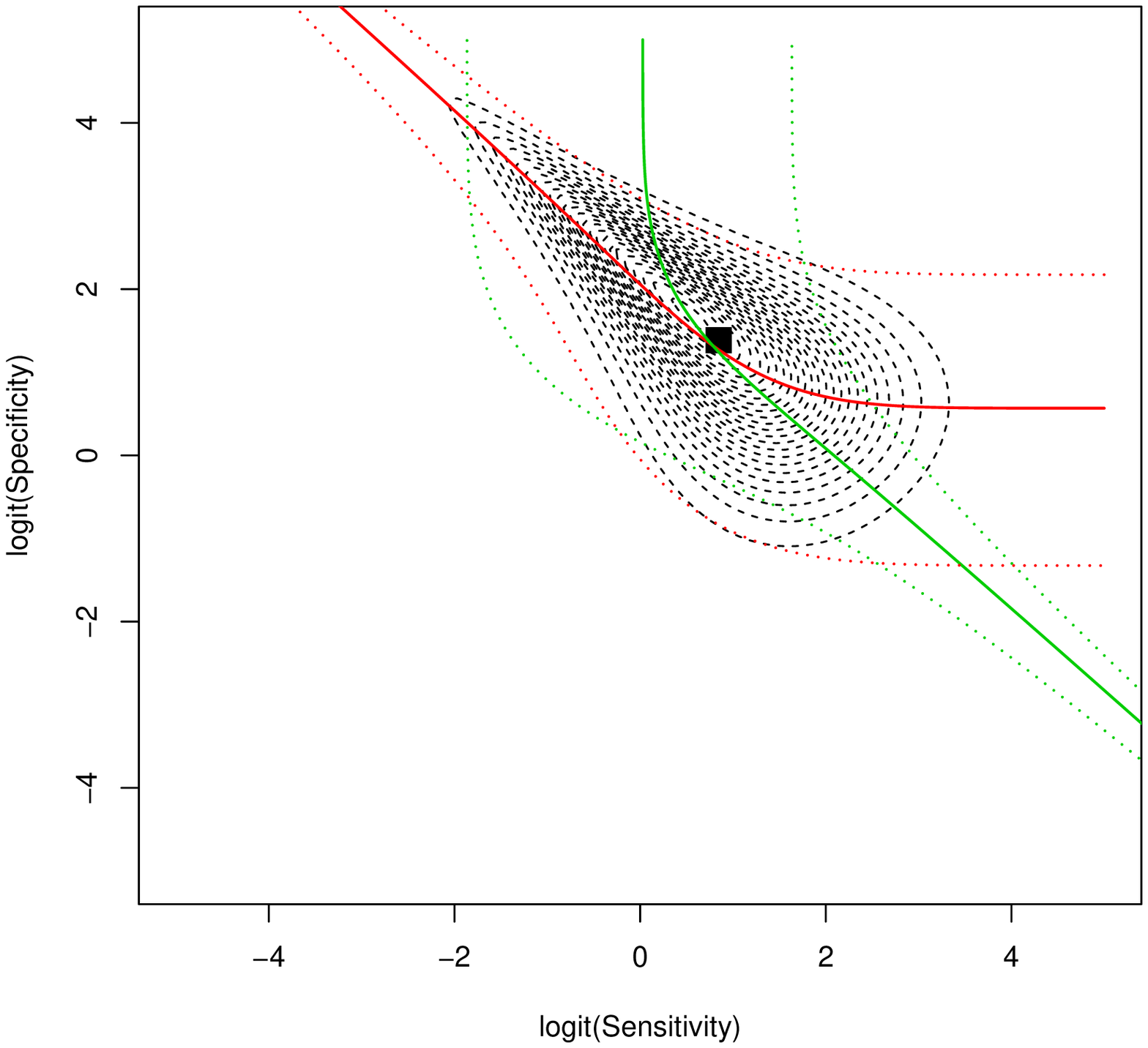}
\\\hline
\end{tabular}
\end{footnotesize}
\caption{\label{ROC-normal}Contour plots and quantile  regression curves  from the copula representation of the random effects distribution with normal  margins and BVN, Frank, and Clayton by 90 and 270  copulas with the same model parameters $\bigl\{\pi_{1}=0.7,\pi_2=0.9,\s_1=2,\s_2=1,\tau=-0.5\bigr\}$. Red and green lines represent the quantile  regression curves $x_1:=\widetilde{x}_1(x_2,q)$ and $x_2:=\widetilde{x}_2(x_1,q)$, respectively; for $q=0.5$ solid lines and for $q\in\{0.01,0.99\}$ dotted lines.}
\end{center}
\vspace{-0.5cm}
\end{figure}

\begin{figure}[!h]
\begin{center}
\begin{footnotesize}
\begin{tabular}{|cc|}

\hline BVN& Frank\\\hline

\includegraphics[width=0.3\textwidth]{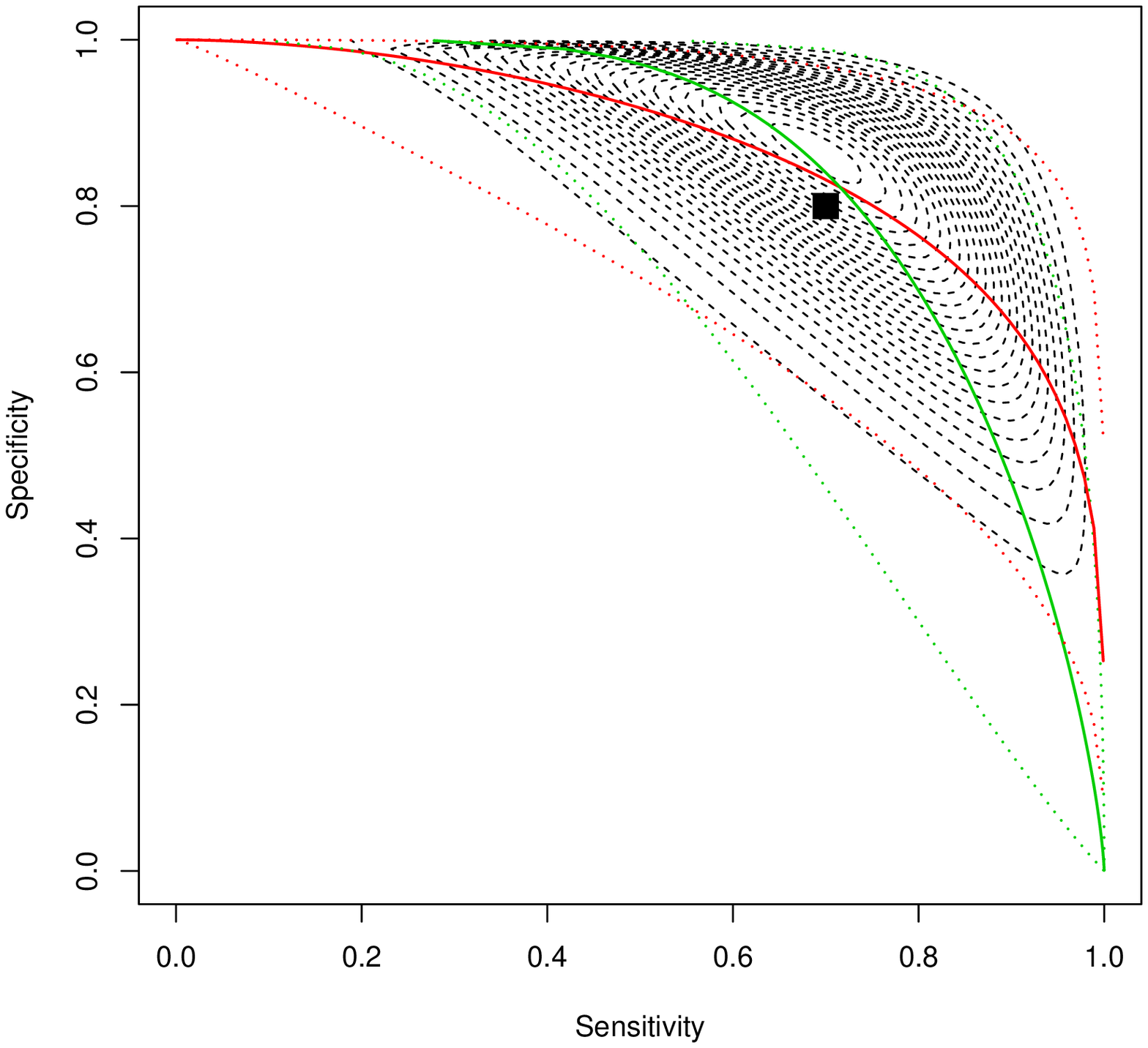}
&

\includegraphics[width=0.3\textwidth]{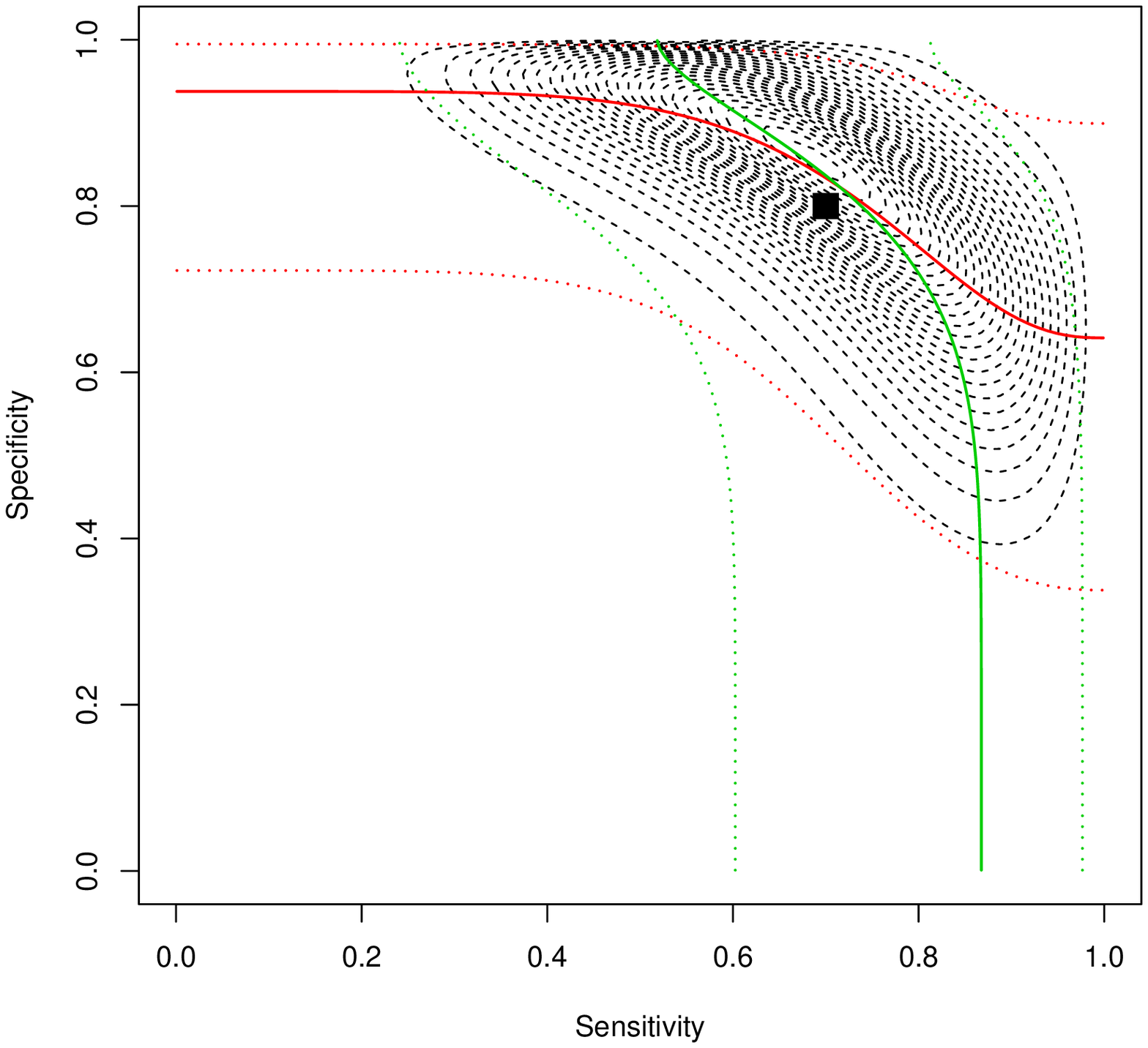}\\\hline
Clayton by 90& Clayton by 270\\\hline
\includegraphics[width=0.3\textwidth]{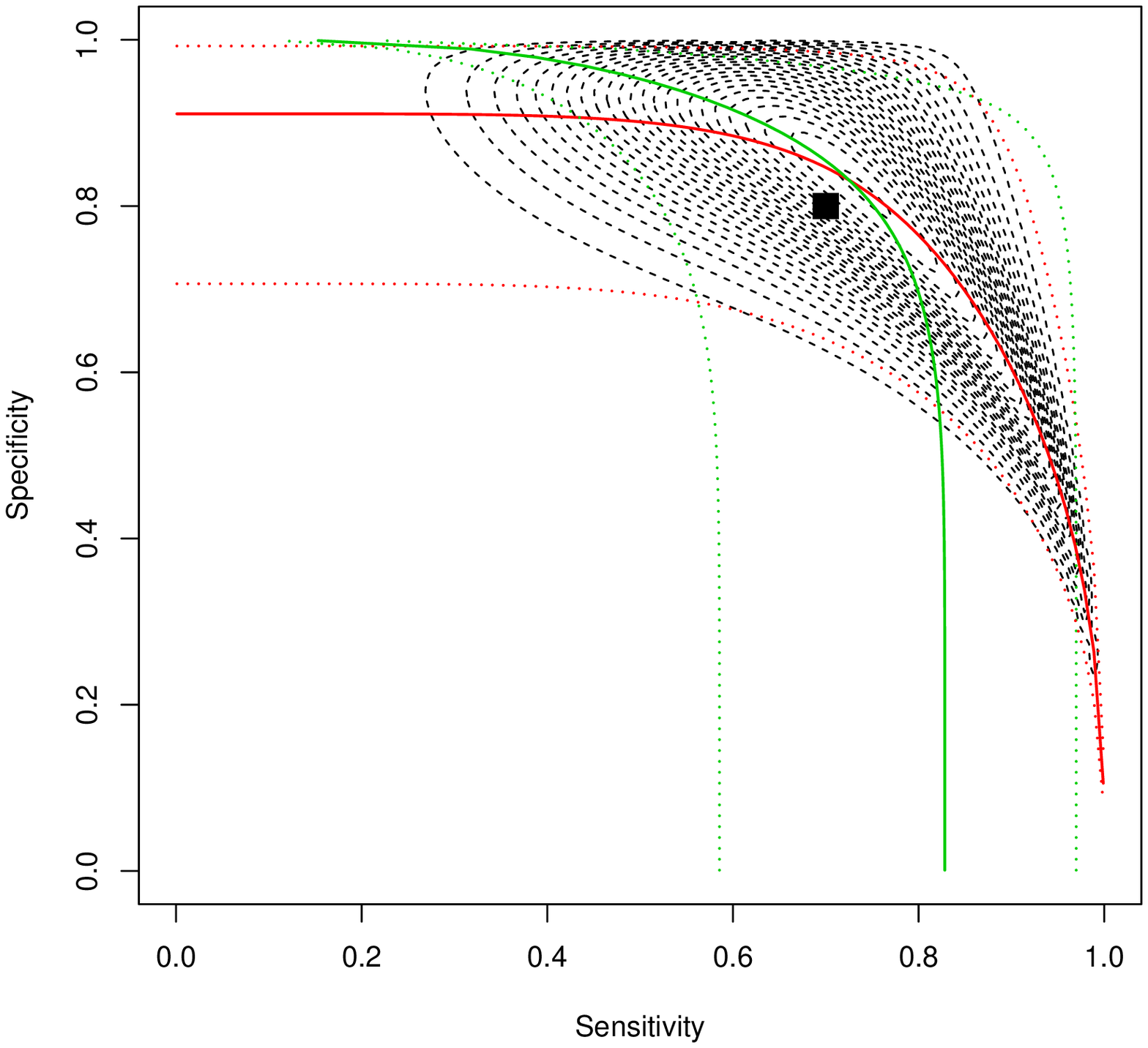}
&
\includegraphics[width=0.3\textwidth]{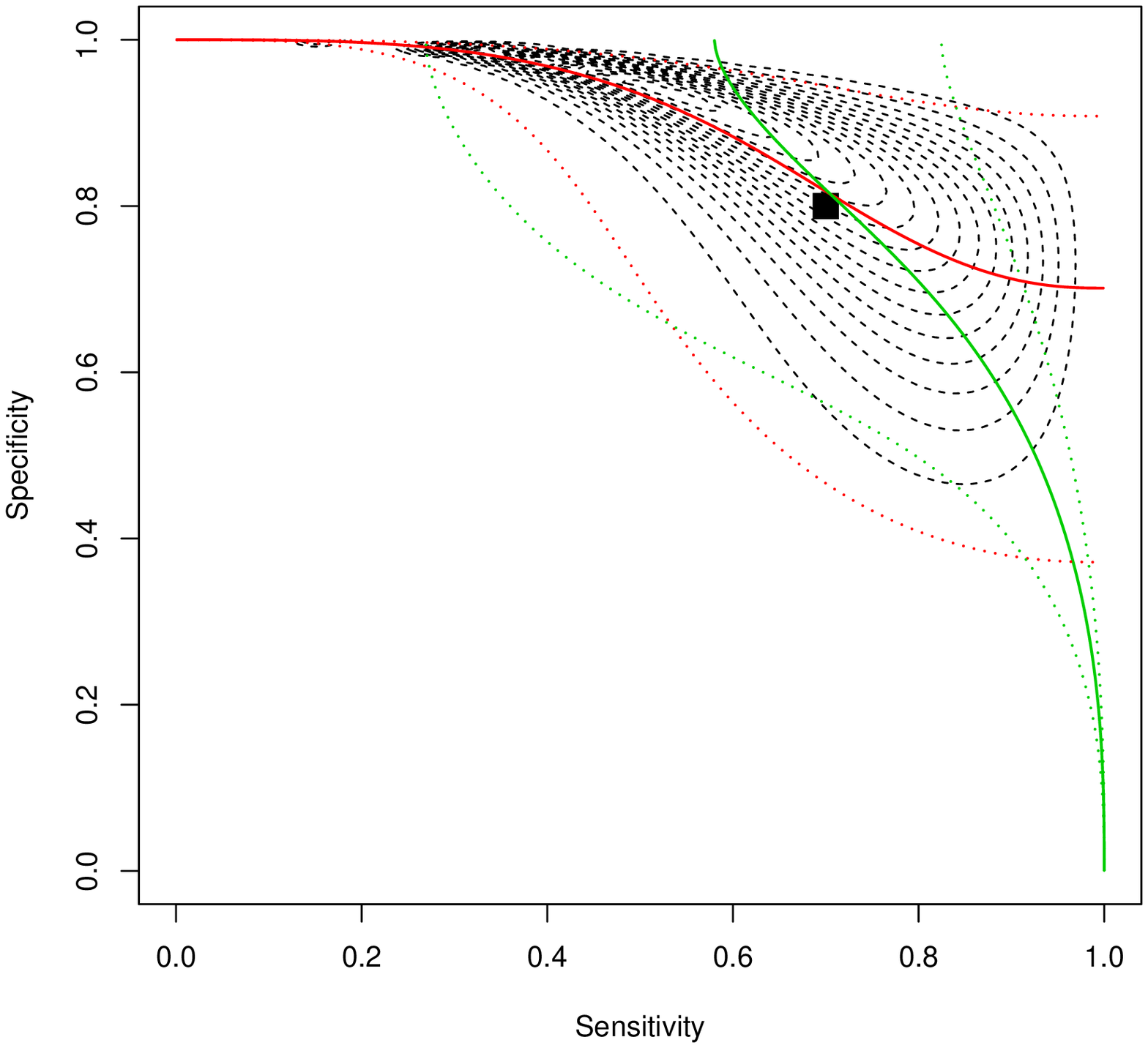}\\\hline
\end{tabular}
\end{footnotesize}
\caption{\label{ROC-beta}Contour plots and quantile  regression curves  from the copula representation of the random effects distribution with beta  margins and BVN, Frank, and Clayton by 90 and 270  copulas with the same model parameters $\bigl\{\pi_{1}=0.7,\pi_2=0.9,\g_1=0.2,\g_2=0.1,\tau=-0.5\bigr\}$. Red and green lines represent the quantile  regression curves $x_1:=\widetilde{x}_1(x_2,q)$ and $x_2:=\widetilde{x}_2(x_1,q)$, respectively; for $q=0.5$ solid lines and for $q\in\{0.01,0.99\}$ dotted lines.}

\end{center}
\vspace{-0.5cm}
\end{figure}

To depict the different shapes of the SROC curves, in Figures \ref{ROC-normal} and \ref{ROC-beta} we plot them from the copula representation of the random effects distribution with normal and beta margins, respectively, and BVN, Frank Clayton by 90 and 270  copulas with the same model parameters $\bigl\{\pi_{1}=0.7,\pi_2=0.9,\s_1=2,\s_2=1,\tau=-0.5\bigr\}$ and $\bigl\{\pi_{1}=0.7,\pi_2=0.9,\g_1=0.2,\g_2=0.1,\tau=-0.5\bigr\}$, respectively.
We convert from $\tau$ to the BVN, Frank and   rotated  Clayton copula parameter $\theta$ via the relations
\begin{equation}\label{tauBVN}
\tau=\frac{2}{\pi}\arcsin(\th),
\end{equation}
\begin{equation}\label{tauFrank}
\tau=\left\{\begin{array}{ccc}
1-4\theta^{-1}-4\theta^{-2}\int_\theta^0\frac{t}{e^t-1}dt &,& \th<0\\
1-4\theta^{-1}+4\theta^{-2}\int^\theta_0\frac{t}{e^t-1}dt &,& \th>0\\
\end{array}\right.,
\end{equation}
\begin{equation}\label{tauCln}
\tau=\left\{\begin{array}{rcl}
\th/(\th+2)&,& \mbox{by 0 or 180 degrees}\\
-\th/(\th+2)&,& \mbox{by 90 or 270 degrees}\\
\end{array}\right.
\end{equation}
in \cite{HultLindskog02}, \cite{genest87}, and \cite{genest&mackay86} respectively.

\section{\label{miss-section}Small-sample efficiency -- Misspecification of the random effects distribution}

An extensive simulation study is conducted
(a) to gauge the small-sample efficiency of the ML and approximation in Kuss {\it et al.} \cite{kuss-etal-2013}'s (hereafter KHS) methods, and
(b) to investigate in detail
the  misspecification of the parametric margin or  family of copulas of the random effects distribution.

To simulate the data we have used   the generation  process in \cite{paul-etal-2010} to get heterogeneous study sizes;  the simulation steps follow:
\begin{enumerate}
\itemsep=10pt
\item Simulate the study size $n$ from a shifted gamma distribution, i.e., $n\sim \mbox{sGamma}(\a=1.2,\b=0.01,\mbox{lag}=30)$ and round off to the nearest integer.
\item Simulate $(u_1,u_2)$ from a parametric family of copulas $C(;\tau)$;  $\tau$ is converted
to BVN, Frank and Clayton rotated by 90/270 dependence parameter $\th$ via the relations  in (\ref{tauBVN}), (\ref{tauFrank}), and (\ref{tauCln}).

\item Convert to beta realizations  via $x_j=F_j^{-1}(u_j,\p_j,\g_j)$ or normal realizations via $x_j=\Phi_j^{-1}\bigl(u_j,l(\pi_j),\s_j\bigr)$ for $j=1,2$; for the latter convert to proportions via $x_j=l^{-1}(x_j)$.
\item Draw the number of diseased $n_{1}$ from a $B(n,0.43)$ distribution.
\item Set  $n_2=n-n_1$, $y_j=n_jx_j$ and then round $y_j$ for $j=1,2$.
\end{enumerate}

We randomly generated $B=10^4$ samples of size {\bf $N = 20, 50$} from  the Clayton rotated by 270 degrees copula mixed model with beta margins.
Table \ref{sim}  contains the
resultant biases, root mean square errors (RMSE), and standard deviations (SD), along with average theoretical variances scaled by $N$, for the MLEs  under different copula choices and margins. The theoretical variances of the MLEs are obtained via the gradients and the Hessian computed
numerically during the maximization process. We also provide biases, RMSEs and SDs for the KHS estimates under the `true' model, i.e., the Clayton rotated by 270 degrees copula mixed model with beta margins.

Conclusions from the values in the table are the following:
\begin{itemize}
\itemsep=8pt
\item ML  with  the the `true' copula mixed  model is highly efficient according to the simulated biases and standard deviations.

\item The MLEs of the meta-analytic parameters are slightly underestimated under copula misspecification. That is, there is some downward bias for these parameters, especially if the ``working" model is not close to Kullback-Liebler distance with the ``true" model, i.e., it is misspecified. For example in the table there is more bias for the Clayton rotated by 90 degrees and Frank copulas since they have different tail dependence from the `true' model, i.e., the rotated Clayton by 270 degrees. An interesting result is that the BVN copula performed rather well under misspecification.

\item The SDs are rather robust to the copula misspecification.

\item The meta-analytic MLEs and SDs are not robust to the margin misspecification, while
the MLE of $\tau$ and  its SD is.

\item The KHS
approximation method yields to biased univariate parameter  estimates.

\item The efficiency of the KHS approximation method is low for the dependence parameter $\tau$.
The parameter $\tau$ is substantially underestimated.
\end{itemize}

The simulation results indicate that the KHS approximation method in \cite{kuss-etal-2013} is an inefficient; hence   flawed method. This was expected, since  theoretically there are serious problems on modelling assumptions under the case of heterogeneous  study sizes. If
the number of true positives and negatives  do not have a common support over different studies,  then one cannot conclude that there is a copula. To make our study complete, we  perform theoretical calculations, similarly to \cite{joe08,nikoloulopoulos13b,
nikoloulopoulos2015},  to
investigate the accuracy of the  approximate copula likelihood method in \cite{kuss-etal-2013} for the special case of a constant size $n$ of groups of diseased and healthy people in the single studies and show whether or not this leads to consistent estimate of the parameters of
the bivariate random effects distribution.
As shown in the Appendix,
the  KHS method leads to asymptotic bias for both the univariate and copula parameters and hence  there is no consistency.
Also given the resultant  substantial asymptotic downward bias for the dependence parameter, the approximation deteriorates, and, hence cannot be used e.g., for prediction purposes via SROC curves. To this end, the KHS approximation method
is not used in the sequel, since its inefficiency has been shown, and, it should be avoided for bivariate meta-analysis of diagnostic test accuracy studies.

The effect of misspecifying the copula choice can be seen  as minimal for both the univariate parameters and Kendall's tau. However, note that (a) the meta-analytic parameters are a univariate inference, and hence it is the univariate marginal distribution that  matters and not the type of the copula, and, (b) as previously emphasized
Kendall's tau only accounts for the dependence dominated by the middle of the data (sensitivities and specificities), and it is expected to be similar amongst  different families of copulas. However, the tail dependence varies, as explained in Section \ref{sec-families}, and is a property to consider when choosing amongst different families of copulas, and, hence affects the shape of SROC curves, i.e., prediction.  SROC  will essentially show the
effect of different  model (random effect distribution) assumptions, since it is an inference that depends on the joint distribution.

\begin{landscape}
\begin{table}[!h]
  \centering
  \caption{\label{sim} Small sample of sizes $N = 20, 50$ simulations ($10^4$ replications) from the Clayton rotated by 270 degrees copula mixed model with beta margins and resultant biases,  root mean square errors (RMSE) and standard deviations (SD), along with the square root of the average theoretical variances ($\sqrt{\bar V}$), scaled by $N$, for the MLEs  under different copula choices and margins. We also provide biases, RMSEs and SDs for the KHS estimates under the `true' model. }
    \begin{footnotesize}

    \begin{tabular}{ccccccccccccc}
    \hline
          &   Margin    & Copula      & \multicolumn{2}{c}{$\pi_1=0.7$} & \multicolumn{2}{c}{$\pi_2=0.9$} & \multicolumn{2}{c}{$\g_1=0.2$} & \multicolumn{2}{c}{$\g_2=0.1$} & \multicolumn{2}{c}{$\tau=-0.5$} \\

          &       &       & $N=20$ & $N=50$ & $N=20$ & $N=50$ & $N=20$ & $N=50$ & $N=20$ & $N=50$ & $N=20$ & $N=50$ \\\hline

    $N$ Bias & Beta  & BVN   & -0.02 & -0.09 & 0.00  & -0.06 & -0.64 & -1.24 & -0.32 & -0.51 & -1.54 & -2.31 \\
          &       & Frank   & -0.23 & -0.72 & 0.10  & 0.27  & -0.59 & -1.08 & -0.39 & -0.72 & -2.01 & -3.77 \\
          &       & Clayton by 90 & -0.11 & -0.31 & -0.02 & -0.13 & -0.50 & -0.81 & -0.20 & -0.14 & -0.23 & 2.74 \\
          &       & Clayton by 270 & -0.01 & -0.08 & 0.03  & 0.04  & -0.65 & -1.25 & -0.40 & -0.78 & -2.30 & -4.57 \\
          & Normal  & BVN   & 0.71  & 1.87  & 0.63  & 1.64  &-&-&-&-&  -1.56 & -2.26 \\
          &       & Frank   & 0.46  & 1.11  & 0.68  & 1.79  &-&-&-&-&  -1.95 & -3.52 \\
          &       & Clayton by 90 & 0.62  & 1.65  & 0.64  & 1.67  &-&-&-&-&  -0.25 & 2.66 \\
          &       & Clayton by 270 & 0.70  & 1.89  & 0.62  & 1.62  &-&-&-&-&  -2.39 & -4.69 \\

          &  KHS     & Clayton by 270 & 1.99  & 5.58  & -0.26 & -0.55 & -0.48 & -1.01 & -1.44 & -3.87 & 7.42  & 20.09 \\\hline
    $N$ SD & Beta  & BVN   & 0.94  & 1.47  & 0.44  & 0.72  & 1.01  & 1.63  & 0.71  & 1.22  & 3.31  & 4.83 \\
          &       & Frank   & 1.00  & 1.58  & 0.42  & 0.68  & 1.01  & 1.62  & 0.67  & 1.13  & 3.36  & 4.84 \\
          &       & Clayton by 90 & 0.97  & 1.50  & 0.46  & 0.78  & 1.10  & 1.80  & 0.87  & 1.52  & 4.83  & 7.15 \\
          &       & Clayton by 270 & 0.94  & 1.47  & 0.43  & 0.70  & 0.97  & 1.56  & 0.64  & 1.07  & 3.12  & 4.54 \\
          & Normal  & BVN   & 1.06  & 1.66  & 0.38  & 0.59  & 4.77  & 7.35  & 5.02  & 7.93  & 3.35  & 4.76 \\
          &       & Frank   & 1.11  & 1.76  & 0.38  & 0.59  & 4.70  & 7.24  & 5.04  & 7.99  & 3.40  & 4.92 \\
          &       & Clayton by 90 & 1.11  & 1.72  & 0.38  & 0.60  & 5.19  & 8.13  & 5.52  & 8.88  & 4.76  & 6.74 \\
          &       & Clayton by 270 & 1.06  & 1.66  & 0.38  & 0.59  & 4.52  & 6.86  & 4.97  & 7.77  & 3.18  & 4.62 \\

          &  KHS     & Clayton by 270 & 1.17  & 1.95  & 0.56  & 0.77  & 1.04  & 1.68  & 0.76  & 0.72  & 1.82  & 1.48 \\\hline
    $N\sqrt{\bar V}$  & Beta  & BVN   & 0.76  & 1.25  & 0.40  & 0.66  & 0.79  & 1.29  & 0.61  & 1.02  & 2.67  & 3.86 \\
          &       & Frank   & 0.77  & 1.27  & 0.37  & 0.59  & 0.82  & 1.33  & 0.58  & 0.95  & 2.62  & 3.78 \\
          &       & Clayton by 90 & 0.77  & 1.25  & 0.39  & 0.65  & 0.83  & 1.34  & 0.64  & 1.10  & 3.00  & 4.11 \\
          &       & Clayton by 270 & 0.75  & 1.22  & 0.38  & 0.60  & 0.77  & 1.25  & 0.56  & 0.90  & 2.56  & 3.72 \\
          & Normal  & BVN   & 0.87  & 1.45  & 0.32  & 0.53  & 3.70  & 5.82  & 4.58  & 7.35  & 2.90  & 3.84 \\
          &       & Frank   & 0.87  & 1.45  & 0.31  & 0.50  & 3.74  & 5.92  & 4.58  & 7.31  & 2.56  & 3.69 \\
          &       & Clayton by 90 & 0.89  & 1.48  & 0.33  & 0.54  & 3.87  & 6.15  & 4.57  & 7.44  & 2.90  & 3.94 \\
          &       & Clayton by 270 & 0.84  & 1.36  & 0.31  & 0.49  & 3.52  & 5.47  & 4.46  & 6.96  & 2.45  & 3.48 \\\hline
    $N$ RMSE & Beta  & BVN   & 0.94  & 1.47  & 0.44  & 0.72  & 1.19  & 2.05  & 0.78  & 1.32  & 3.65  & 5.35 \\
          &       & Frank   & 1.02  & 1.74  & 0.43  & 0.73  & 1.17  & 1.95  & 0.77  & 1.34  & 3.92  & 6.13 \\
          &       & Clayton by 90 & 0.97  & 1.53  & 0.46  & 0.80  & 1.21  & 1.97  & 0.89  & 1.53  & 4.84  & 7.65 \\
          &       & Clayton by 270 & 0.94  & 1.47  & 0.43  & 0.70  & 1.17  & 2.00  & 0.75  & 1.33  & 3.87  & 6.44 \\
          & Normal  & BVN   & 1.27  & 2.50  & 0.74  & 1.74  &-&-&-&-& 3.69  & 5.27 \\
          &       & Frank   & 1.20  & 2.08  & 0.78  & 1.88  &-&-&-&-& 3.92  & 6.05 \\
          &       & Clayton by 90 & 1.27  & 2.39  & 0.75  & 1.77  &-&-&-&-& 4.77  & 7.24 \\
          &       & Clayton by 270 & 1.27  & 2.52  & 0.73  & 1.73  &-&-&-&-& 3.98  & 6.58 \\

          & KHS      & Clayton by 270 & 2.31  & 5.91  & 0.62  & 0.94  & 1.15  & 1.96  & 1.63  & 3.93  & 7.64  & 20.14 \\
    \hline
    \end{tabular}
  \end{footnotesize}
\end{table}
\end{landscape}

\section{\label{vuong-sec}Vuong's test for model comparison}
In this section we provide a methodology for the comparison of non-nested models.
It would  be used as a tool to show if the copula mixed model provides better fit than the standard GLMM. We will call a test proposed by Vuong     \cite{vuong1989}. The Vuong's test is the sample version of the difference in Kullback-Leibler divergence between two models and can be used to differentiate two  parametric models which could be non-nested.This test has been used extensively in the copula literature to compare copula models, see e.g., \cite{belgorodski10,Brechmann-Czado-Aas-2012,joe2014}

Assume that we have Models 1 and 2 with parametric densities $f^{(1)}$ and  $f^{(2)}$ respectively.
We can compare
$$\Delta_{1f^\maltese}=N^{-1}\Bigl[\sum_i\{E_{f^\maltese}[\log f^\maltese(Y_1,Y_2)]-E_{f^\maltese}[\log f^{(1)}(Y_1,Y_2;\thbf^{(1)})]\}\Bigr],$$
and
$$\Delta_{2f^\maltese}=N^{-1}\Bigl[\sum_i\{E_{f^\maltese}[\log f^\maltese(Y_1,Y_2)]-E_{f^\maltese}[\log f^{(2)}(Y_1,Y_2;\thbf^{(2)})]\}\Bigr],$$
where $\thbf^{(1)},\thbf^{(2)}$ are the parameters in Models 1 and 2 respectively that lead to the closest Kullback-Leibler divergence to the true $f^\maltese$; equivalently they are the limits in probability of the MLEs based on models 1 and 2 respectively. Model 1 is closer to the true $f^\maltese$, i.e., is the better fitting model if $\Delta=\Delta_{1f^\maltese}-\Delta_{2f^\maltese}<0$, and Model 2 is the better fitting model if $\Delta>0$. The sample version of $\Delta$ with MLEs $\hat\thbf^{(1)},\hat\thbf^{(2)}$ is
$$\bar D=\sum_{i=1}^N D_i/N,$$
where $D_i=\log\left[\frac{f^{(2)}\left(Y_1,Y_2;\hat\thbf^{(2)}\right)}{f^{(1)}\left(Y_1,Y_2;\hat\thbf^{(1)}\right)}\right]$.
Vuong \cite{vuong1989}
has shown that asymptotically
$$\sqrt{N}\bar D/s\sim N(0,1),$$
where $s^2=\frac{1}{N-1}\sum_{i=1}^N(D_i-\bar D)^2$.

\section{\label{sec-appl}Illustrations}
We illustrate the use of the copula mixed model  by re-analysing
the data of two published meta-analysis    \cite{Scheidler-etal-1997,glas-etal-2003}. These data have been frequently used as an example for
methodological papers on meta-analysis of diagnostic accuracy studies  \cite{Reitsma-etal-2005,Chu&Cole2006,RutterGatsonis2001,Riley-etal-2007,Harbord&Whiting2009,kuss-etal-2013}.

We fit the copula mixed model for all different choices of parametric families of copulas and margins.
To make it easier
to compare strengths of dependence, we convert the copula parameters to Kendall's $\tau$'s
  via the relations in (\ref{tauBVN}), (\ref{tauFrank}), and (\ref{tauCln})  for BVN, Frank and rotated Clayton copulas, respectively.
Since the number of parameters is the same between the models, we  use the  log-likelihood at  estimates as a rough diagnostic measure for goodness of fit between the  models. We further compute the Vuong's tests with Model 1 being the BVN copula mixed model with normal margins, i.e., the standard GLMM, to reveal if any other copula mixed model provides better fit than the standard GLMM.

Finally, we demonstrate SROC curves and summary operating points (a pair of average sensitivity and specificity) with a confidence region and a predictive region as deduced in Section \ref{sec-SROC}.

\subsection{The telomerase and computed tomography data}
In Glas {\it et al.} \cite{glas-etal-2003} the telomerase marker for the diagnosis of
bladder cancer is evaluated using 10 studies.  The  size in each study ranges from 35 to 195. The interest was to define if this non-invasive and cheap marker could replace the standard of cystoscopy or histopathology. Riley {\it et al.} \cite{Riley-etal-2007} applied the GLMM  with different starting values and all produced a between-study correlation estimate of $-1$ but with different meta-analytic parameter point estimates and standard errors. Clearly at this example, it is not possible to estimate the correlation between the logit sensitivity and specificity, and the maximum likelihood estimator should truncate the correlation to the left boundary of its parameter space, i.e., $-1$. In \cite{kuss-etal-2013} it is acknowledged that the copula model for observed discrete variables which have beta-binomial margins yields sensible estimates for the dependence parameter and its standard error. This result is in error and  due to the fact that the KHS method underestimates the dependence parameter as emphasized in Section \ref{miss-section} and shown in the Appendix for $\rho=-1$.

Fitting the copula mixed model for all different choices of parametric families of copulas and margins,  the resultant estimate of the  dependence parameter was close to the boundary of its parameter space. If the dependence parameter estimate is so large (on absolute value), the copula should be set to countermonotonic (Fr\'echet lower bound), and, then optimize over the remaining (univariate) parameters. With other words there exists negative perfect dependence, and thus there is  only one copula: the countermonotonic copula. This is a limiting case for all the parametric families of copulas, listed in Section \ref{sec-families}, when the dependence parameter is fixed to the left boundary of its parameter space.

This was also the case for the analysis of the data on 17 studies of computed tomography (CT) for the diagnosis of lymph node metastasis in women with cervical cancer, one of three imaging techniques in the meta-analysis in \cite{Scheidler-etal-1997}. The  size in each study ranges from 20 to 253. Diagnosis of metastatic disease by CT relies on nodal enlargement.

\bigskip 

\begin{table}[!h]
  \centering
  \caption{\label{telomerase-res}Maximised  log-likelihoods, estimates  and standard errors (SE), along with the Vuong's statistics and $p$-values for the telomerase and computed tomography data.}
\begin{footnotesize}
    \begin{tabular}{cccccc|cccccc}
    \hline
    \multicolumn{5}{c}{Telomerase}        &       & \multicolumn{6}{c}{Computed Tomography} \\
    \hline
    \multicolumn{3}{c}{Normal margins} & \multicolumn{3}{c}{Beta margins} & \multicolumn{3}{c}{Normal margins} & \multicolumn{3}{c}{Beta margins} \\\hline
          & Est. & SE    &       & Est. & SE    &       & Est. & SE    &       & Est. & SE \\\hline
    $\pi_1$ & 0.77  & 0.03  & $\pi_1$ & 0.76  & 0.03  & \multicolumn{1}{c}{$\pi_1$} & 0.46  & 0.07  & $\pi_1$ & 0.46  & 0.06 \\
    $\pi_2$ & 0.91  & 0.05  & $\pi_2$ & 0.81  & 0.06  & \multicolumn{1}{c}{$\pi_2$} & 0.93  & 0.01  & $\pi_2$ & 0.92  & 0.01 \\
    $\s_1$ & 0.43  & 0.13  & $\g_1$ & 0.03  & 0.02  & \multicolumn{1}{c}{$\s_1$} & 1.00  & 0.27  & $\g_1$ & 0.17  & 0.07 \\
    $\s_2$ & 1.83  & 0.40  & $\g_2$ & 0.28  & 0.10  & \multicolumn{1}{c}{$\s_2$} & 0.60  & 0.23  & $\g_2$ & 0.02  & 0.02 \\\hline
    $\log L$ & \multicolumn{2}{c}{-50.37} & $\log L$ & \multicolumn{2}{c}{-51.14} & \multicolumn{1}{|c}{$\log L$} & \multicolumn{2}{c}{-69.37} & $\log L$ & \multicolumn{2}{c}{-69.58}  \\\hline
    \multicolumn{6}{c}{Vuong's test}              & \multicolumn{6}{c}{Vuong's test} \\\hline
   $\sqrt{N}\bar D/s$       & \multicolumn{2}{c}{-} &    $\sqrt{N}\bar D/s$   & \multicolumn{2}{c|}{-1.580} & \multicolumn{1}{c}{} & \multicolumn{2}{c}{-} &       & \multicolumn{2}{c}{-1.416} \\
   $p$-value       & \multicolumn{2}{c}{-} &       $p$-value & \multicolumn{2}{c|}{0.114} & \multicolumn{1}{c}{} & \multicolumn{2}{c}{-} &       & \multicolumn{2}{c}{0.157} \\
    \hline
    \end{tabular}
    \end{footnotesize}
\end{table}

\bigskip

Table \ref{telomerase-res}
gives the estimated  univariate parameters, standard errors, and log-likelihoods for both normal and beta margins for both datasets.
For telomerase data, both models agree on the estimated sensitivity $\hat\pi_1$ but the estimate of specificity $\hat\pi_2$ is larger under the standard GLMM.   The log-likelihood  is $-50.37$ for normal margins and $-51.14$ for beta margins, and thus a normal margin seems to be a better fit for the data. Furthermore, according to the Vuong's test the copula mixed model with normal margins (i.e., the standard GLMM) provides marginally better fit ($p$-value$=0.114$).
For computed tomography data, both models agree on the estimated sensitivity $\hat\pi_1$ and  specificity $\hat\pi_2$.  The log-likelihood  is $-69.37$ for normal margins and $-69.58$ for beta margins, and thus a normal margin seems to be a better fit for the data. However,  according to the Vuong's test the copula mixed model with normal margins (i.e., the standard GLMM) does not provide statistical significant  better fit ($p$-value$=0.157$).

\begin{figure}[!h]
\begin{center}
\begin{footnotesize}
\begin{tabular}{|cc|}
\hline
Telomerase & Computed Tomography\\
 \hline
\includegraphics[width=0.3\textwidth]{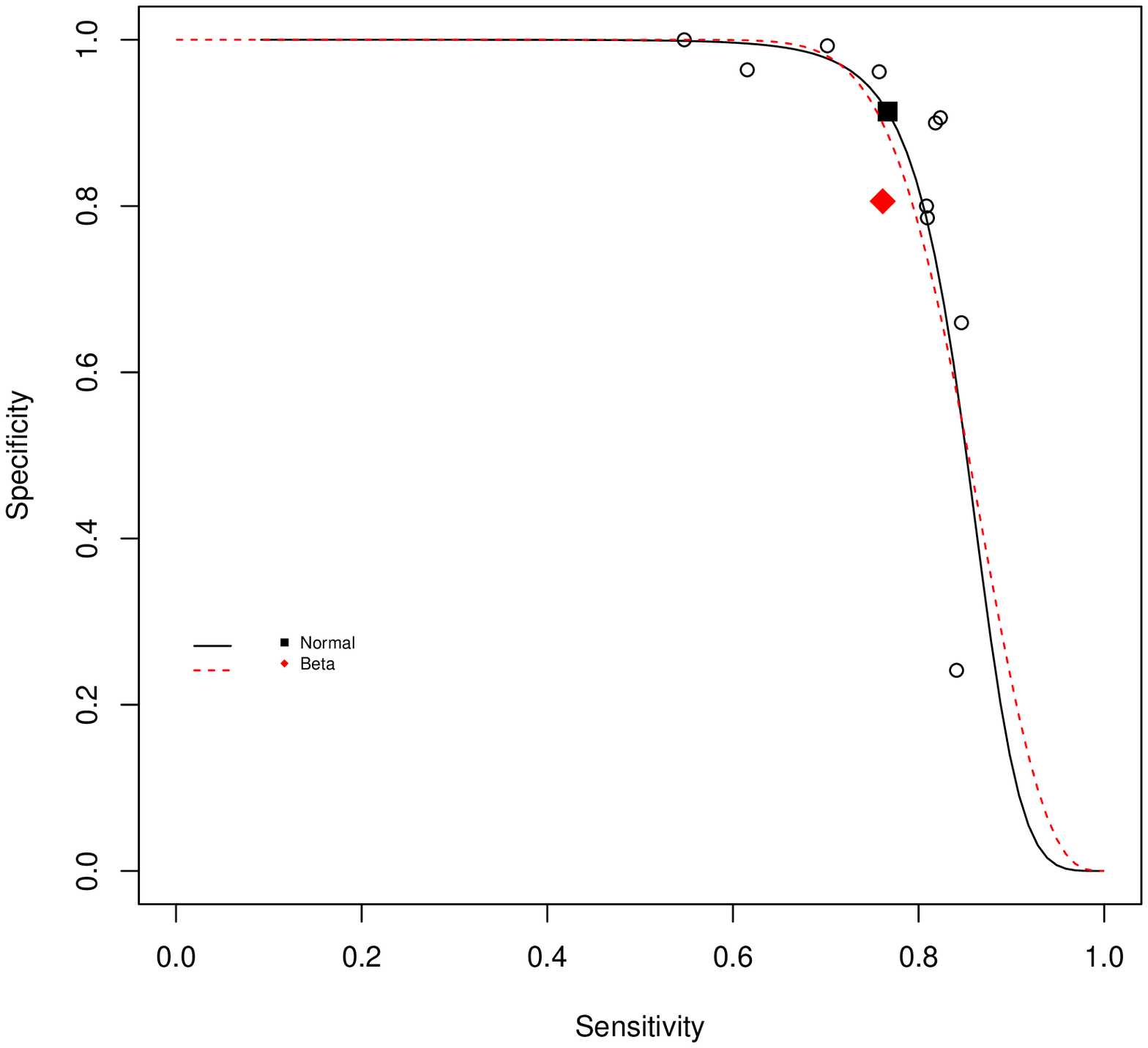}
&
\includegraphics[width=0.3\textwidth]{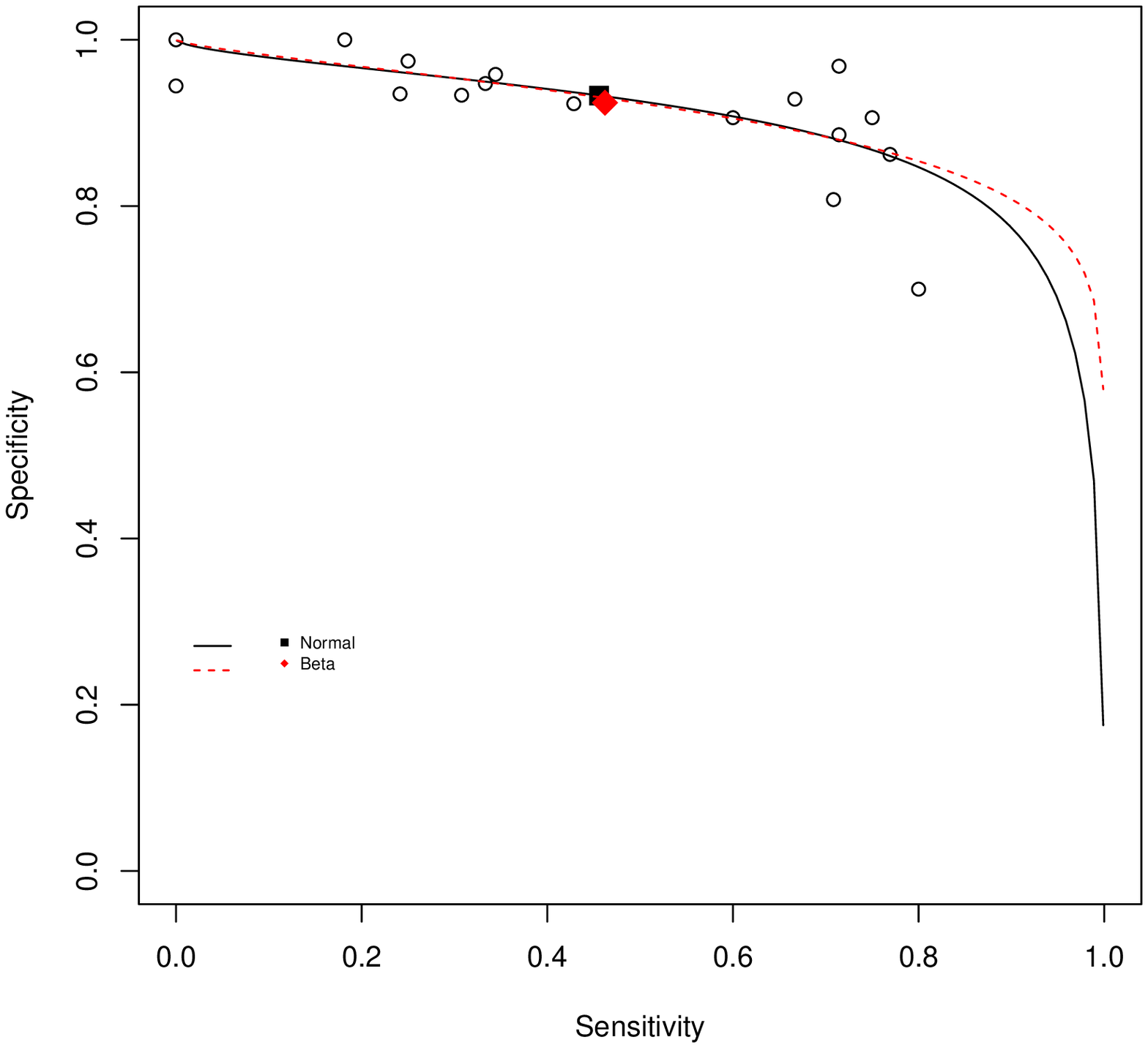}\\\hline
\end{tabular}
\end{footnotesize}
\caption{\label{SROC-tel}SROC curves   from the countermonotonic copula representation of the random effects distribution with normal  margins (black line) and beta (red line) margins for the telomerase and computed tomography data.}
\end{center}
\vspace{-0.5cm}
\end{figure}

Finally, figure \ref{SROC-tel} also shows the SROC curves  for both datasets and the visual fit is consistent with the model fitting and comparison in Table \ref{telomerase-res}. Note in passing since we are dealing with the countermonotonic copula all the quantile regression curves almost coincide, and hence we just depict one median regression curve for each model.

\pagebreak

\subsection{The lymphangiography data}
In this section we apply the copula mixed models  to data on 17 studies of lymphangiography for the diagnosis of lymph node metastasis in women with cervical cancer, one of three imaging techniques in the meta-analysis in \cite{Scheidler-etal-1997}. The  size in each study ranges from 21 to 300. Diagnosis of metastatic disease by lymphangiography is based on the presence of nodal-filling defects.

\begin{table}[!h]
\caption{\label{lag-res}Maximised  log-likelihoods, estimates and standard errors (SE), along with the Vuong's statistics and $p$-values for the lymphangiography data.}
\centering
\begin{footnotesize}
\begin{tabular}{c|cc|cc|cc|cc}
\hline
                                                                               \multicolumn{ 9}{c}{Normal margins} \\
\hline
           & \multicolumn{ 2}{c|}{BVN} & \multicolumn{ 2}{c|}{Frank} & \multicolumn{ 2}{c|}{Clayton by 180} & \multicolumn{ 2}{c}{Clayton by 270} \\

           &   Estimate &         SE &   Estimate &         SE &   Estimate &         SE &   Estimate &         SE \\
\hline
  $\pi_1$ &       0.67 &       0.03 &       0.68 &       0.03 &       0.67 &  0.03  &       0.67 &       0.03 \\

  $\pi_2$ &       0.84 &       0.03 &       0.84 &       0.03 &       0.84 &       0.03 &       0.84 &       0.04 \\

   $\s_1$ &       0.35 &       0.19 &       0.36 &       0.18 &       0.34 &       0.18 &       0.34 &       0.19 \\

   $\s_2$ &       0.91 &       0.22 &       0.91 &       0.22 &       0.91 &       0.22 &       0.90 &       0.22 \\

    $\tau$ &       0.16 &       0.29 &       0.18 &       0.28   &       0.14 &       0.21 &      0.19 &  0.29\\
\hline
   $\log L$ & \multicolumn{ 2}{c}{-91.38} & \multicolumn{ 2}{c}{-91.32} & \multicolumn{ 2}{c}{-91.32} & \multicolumn{ 2}{c}{-91.15} \\
\hline
\multicolumn{9}{c}{Vuong's test }\\\hline
   $\sqrt{N}\bar D/s$        &   \multicolumn{ 2}{c}{-} & \multicolumn{ 2}{c}{0.523} &  \multicolumn{ 2}{c}{0.274}  &\multicolumn{ 2}{c}{1.280} \\

      $p$-value     &   \multicolumn{ 2}{c}{-} & \multicolumn{ 2}{c}{0.601}  & \multicolumn{ 2}{c}{0.784}&\multicolumn{ 2}{c}{0.201}  \\
\hline
                                                                                 \multicolumn{ 9}{c}{Beta margins} \\
\hline
           & \multicolumn{ 2}{c|}{BVN } & \multicolumn{ 2}{c|}{Frank} & \multicolumn{ 2}{c|}{Clayton by 180} & \multicolumn{ 2}{c}{Clayton by 270} \\
                      &   Estimate &         SE &   Estimate &         SE &   Estimate &         SE &   Estimate &         SE \\
\hline
  $\pi_1$ &       0.67 &       0.03 &       0.67 &       0.03 &       0.67 &       0.03 &       0.67 &       0.03 \\

  $\pi_2$ &       0.81 &       0.03 &       0.81 &       0.03 &       0.81 &       0.03 &       0.81 &       0.03 \\

   $\g_1$ &       0.03 &       0.03 &       0.03 &       0.03 &       0.02&       0.03 &       0.02 &       0.03 \\

   $\g_2$ &       0.09 &       0.04 &       0.09 &       0.04&       0.10 &       0.04 &       0.09 &       0.04  \\

    $\tau$ &       0.15 &       0.30 &       0.18 &       0.32&       0.16 &       0.40 &      0.19 &       0.28  \\
\hline
   $\log L$ & \multicolumn{ 2}{c}{-90.67} & \multicolumn{ 2}{c}{-90.61} & \multicolumn{ 2}{c}{-90.60} & \multicolumn{ 2}{c}{-90.44} \\
\hline
\multicolumn{9}{c}{Vuong's test}\\\hline
   $\sqrt{N}\bar D/s$        & \multicolumn{ 2}{c}{1.668} & \multicolumn{ 2}{c}{1.798} &
   \multicolumn{ 2}{c}{1.877}& \multicolumn{ 2}{c}{2.248}   \\

   $p$-value        & \multicolumn{ 2}{c}{0.095} & \multicolumn{2}{c}{0.072} & \multicolumn{ 2}{c}{0.061} & \multicolumn{ 2}{c}{0.025} \\
\hline
\end{tabular}
\end{footnotesize}
\end{table}

\begin{figure}[!h]
\begin{center}
\begin{footnotesize}
\begin{tabular}{|cc|}

\hline BVN& Frank\\\hline

\includegraphics[width=0.3\textwidth]{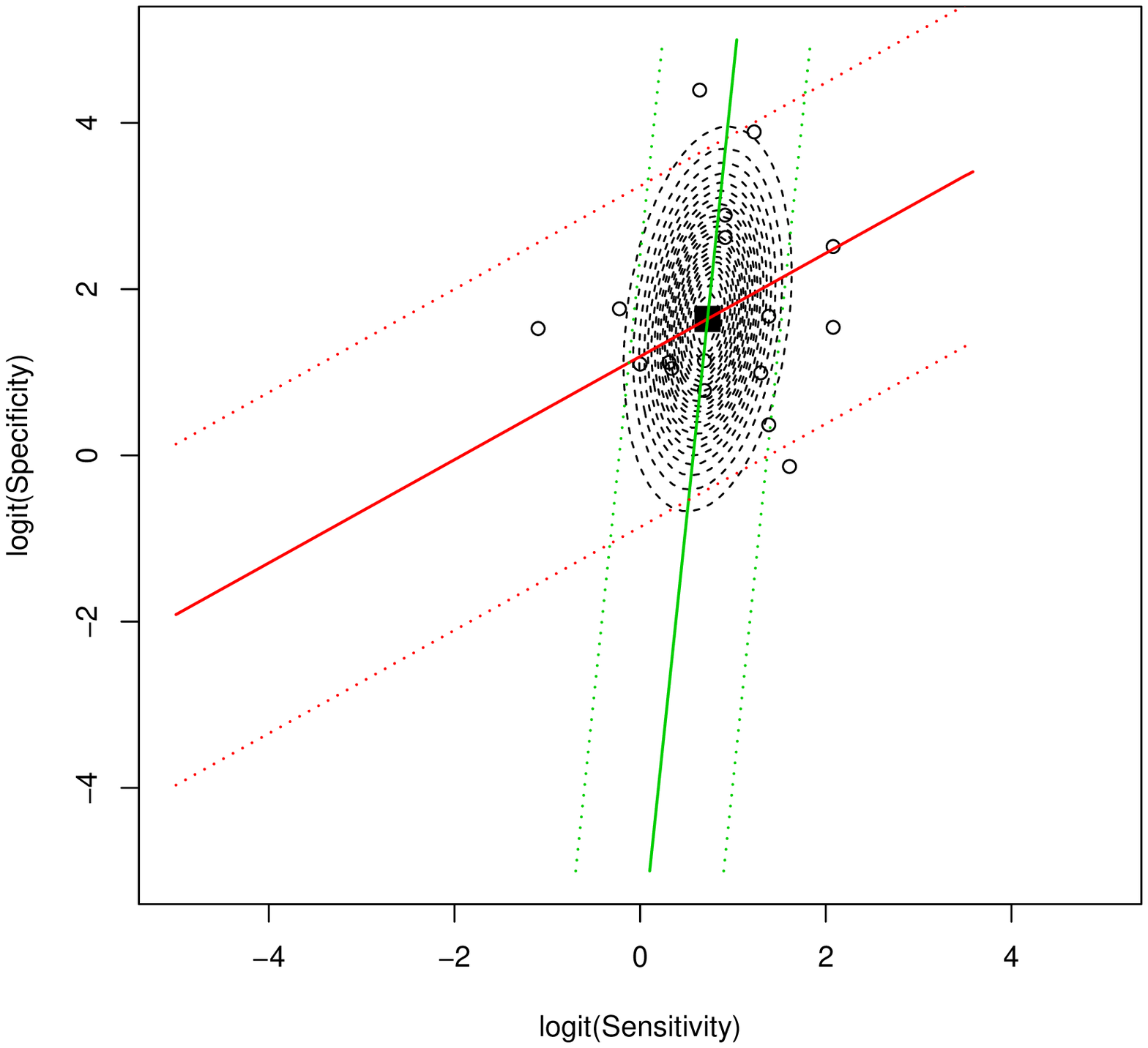}
&

\includegraphics[width=0.3\textwidth]{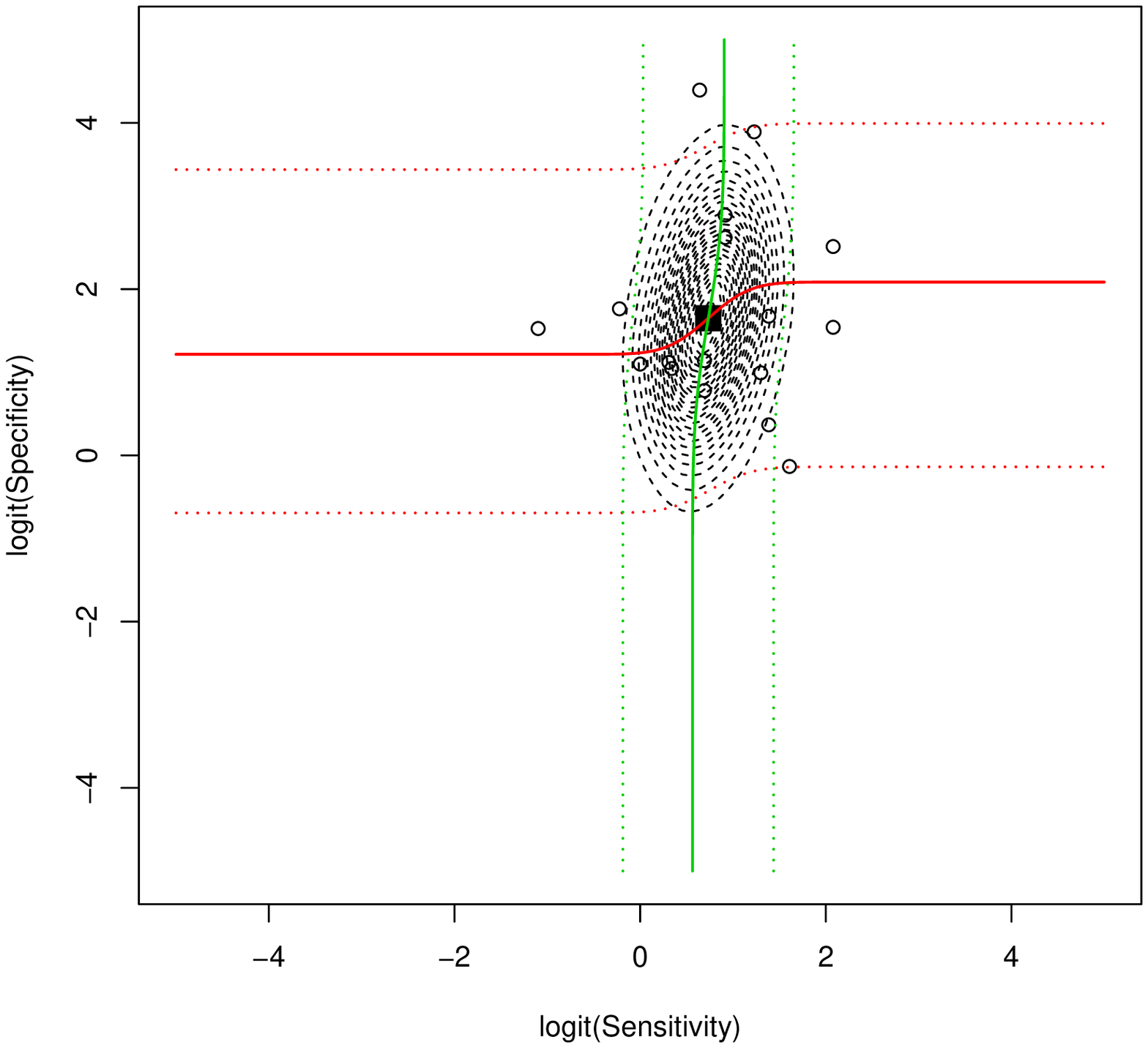}\\\hline
Clayton by 180& Clayton by 270\\\hline
\includegraphics[width=0.3\textwidth]{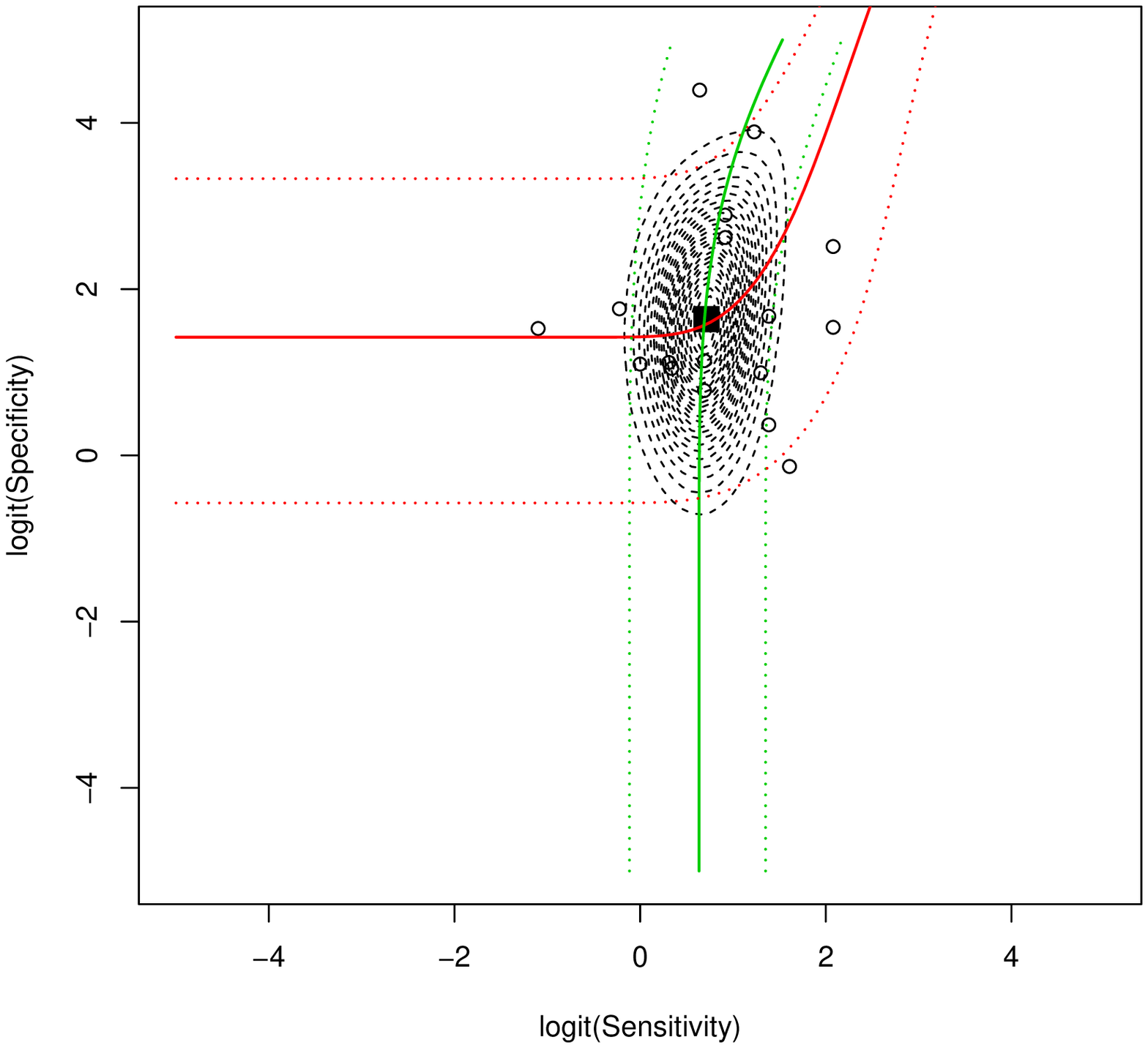}
&
\includegraphics[width=0.3\textwidth]{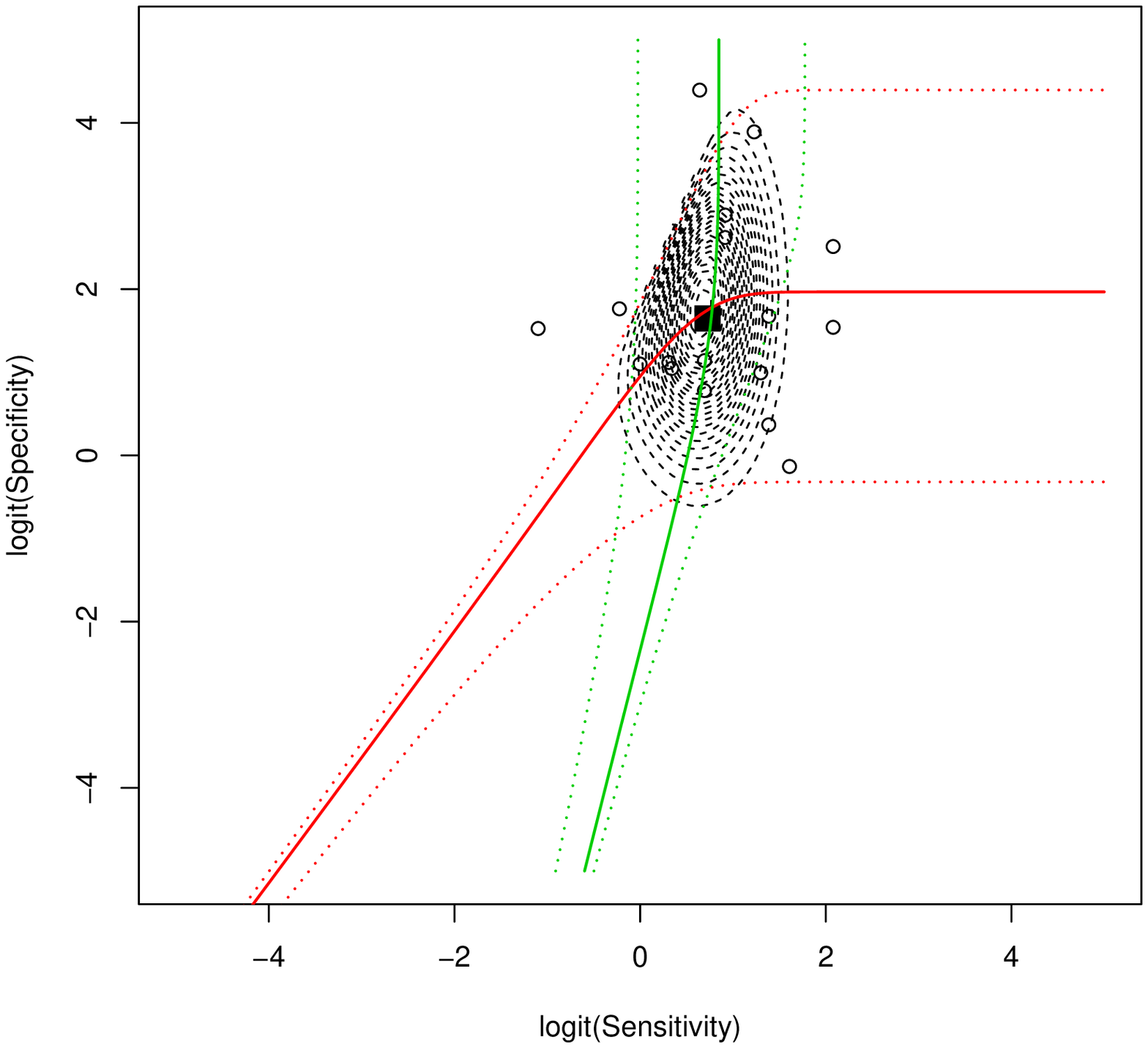}\\\hline
\end{tabular}
\end{footnotesize}
\caption{\label{SROC-LAG-normal}Contour plots and quantile  regression curves  from the copula representation of the random effects distribution with normal  margins and BVN, Frank, and Clayton by 180 and 270  copulas for the lymphangiography data. Red and green lines represent the quantile  regression curves $x_1:=\widetilde{x}_1(x_2,q)$ and $x_2:=\widetilde{x}_2(x_1,q)$, respectively; for $q=0.5$ solid lines and for $q\in\{0.01,0.99\}$ dotted lines.}
\end{center}
\vspace{-0.5cm}
\end{figure}

\begin{figure}[!h]
\begin{center}
\begin{footnotesize}
\begin{tabular}{|cc|}

\hline BVN& Frank\\\hline

\includegraphics[width=0.3\textwidth]{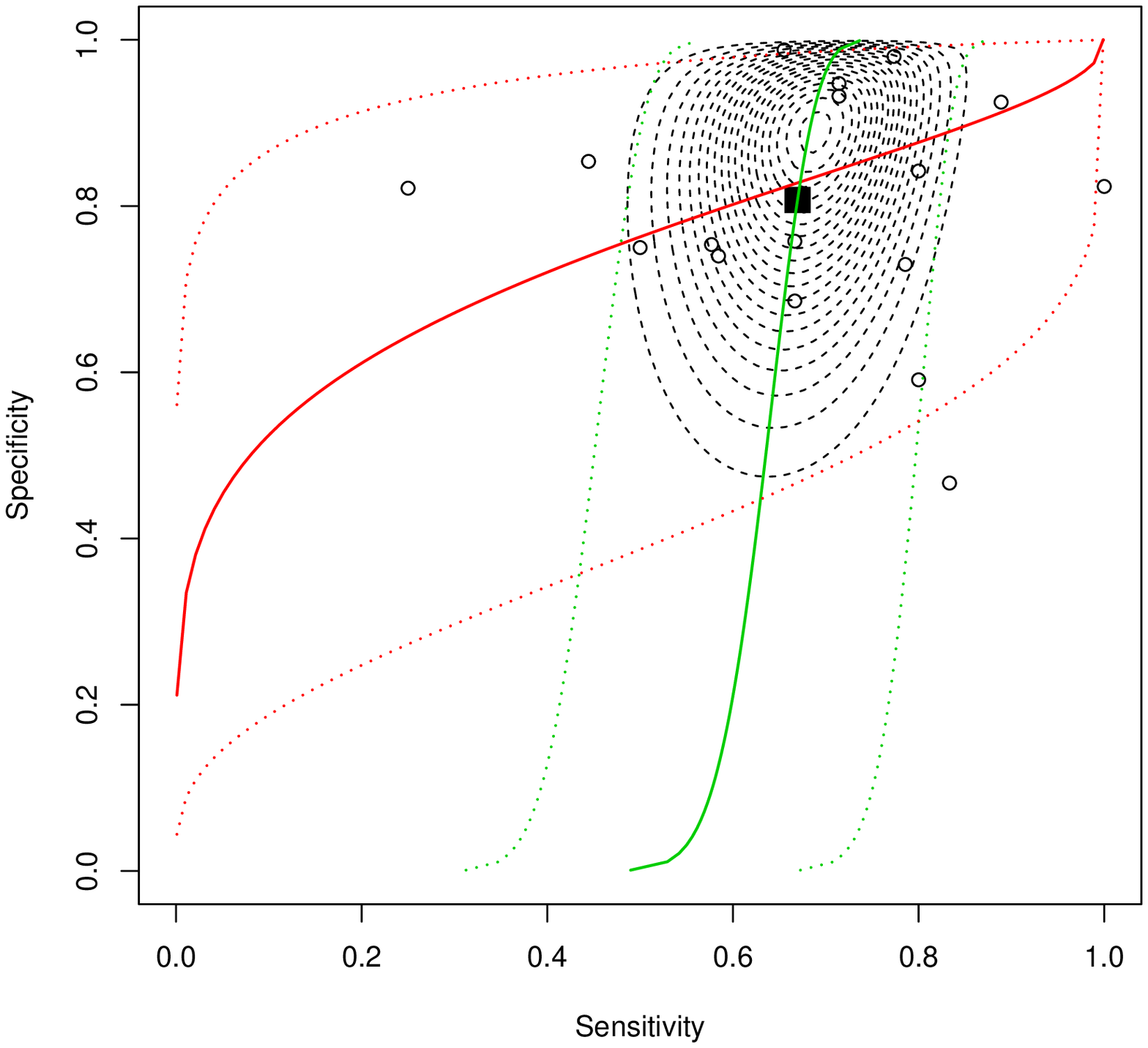}
&

\includegraphics[width=0.3\textwidth]{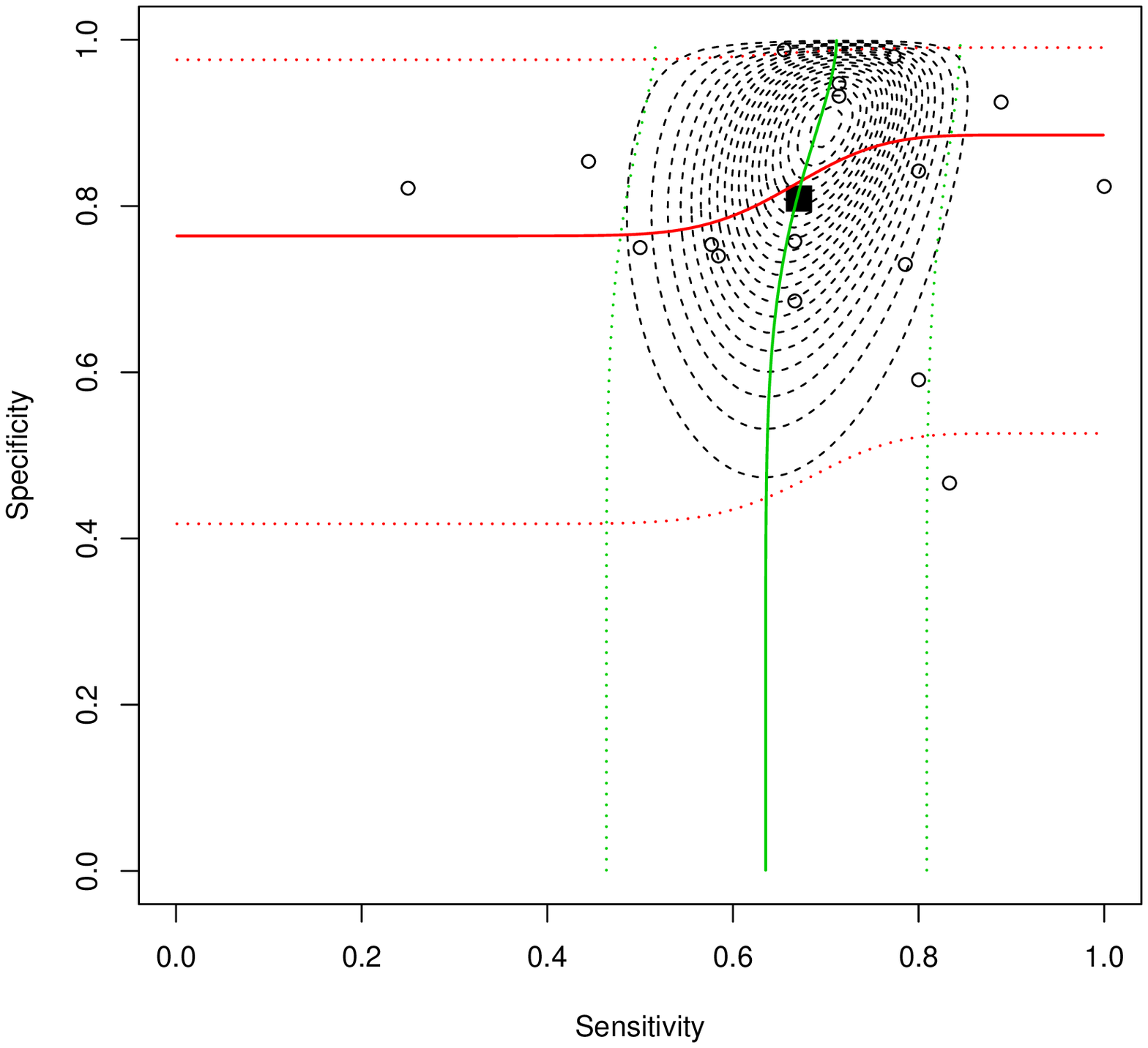}\\\hline
Clayton by 180& Clayton by 270\\\hline
\includegraphics[width=0.3\textwidth]{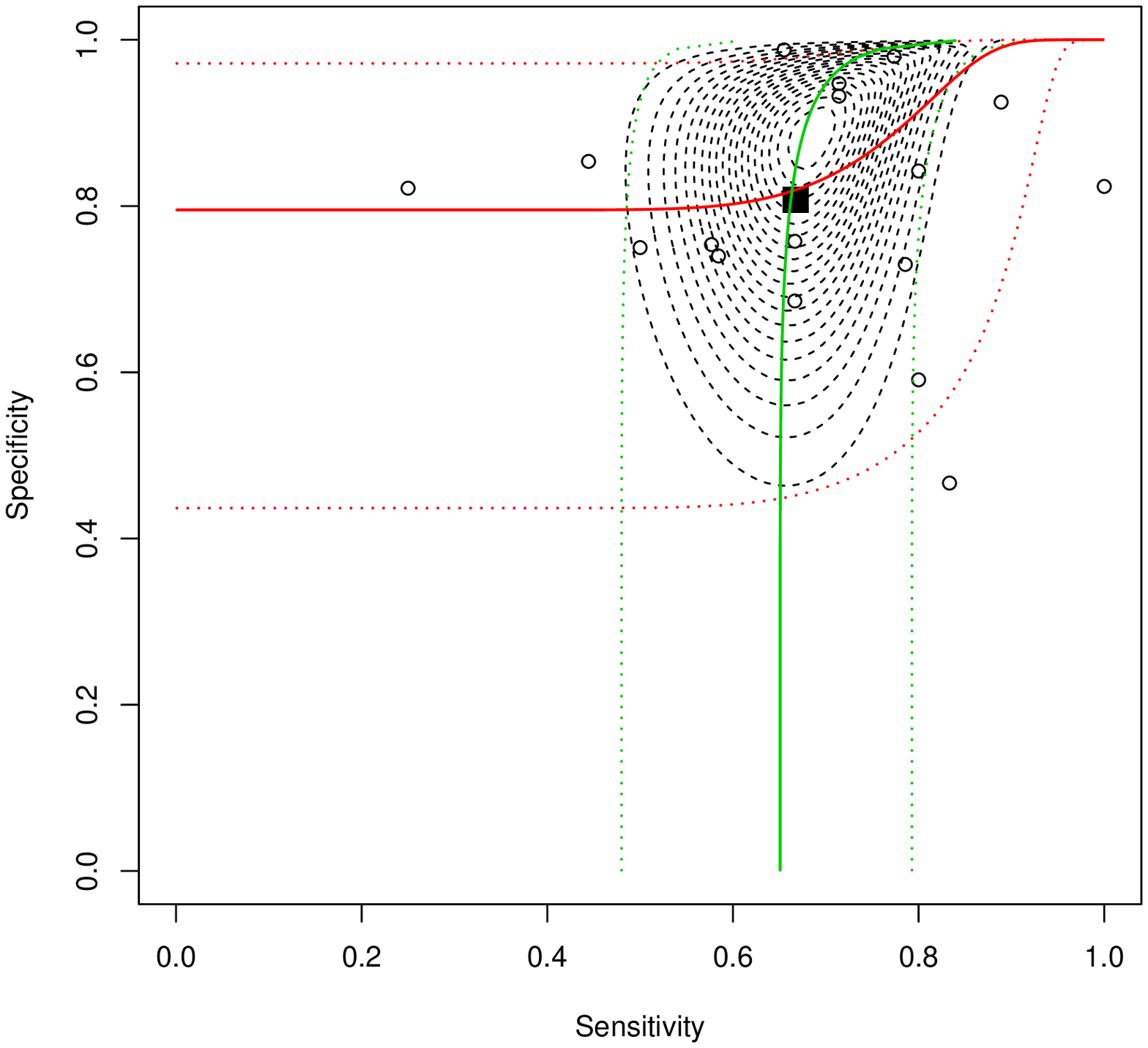}
&
\includegraphics[width=0.3\textwidth]{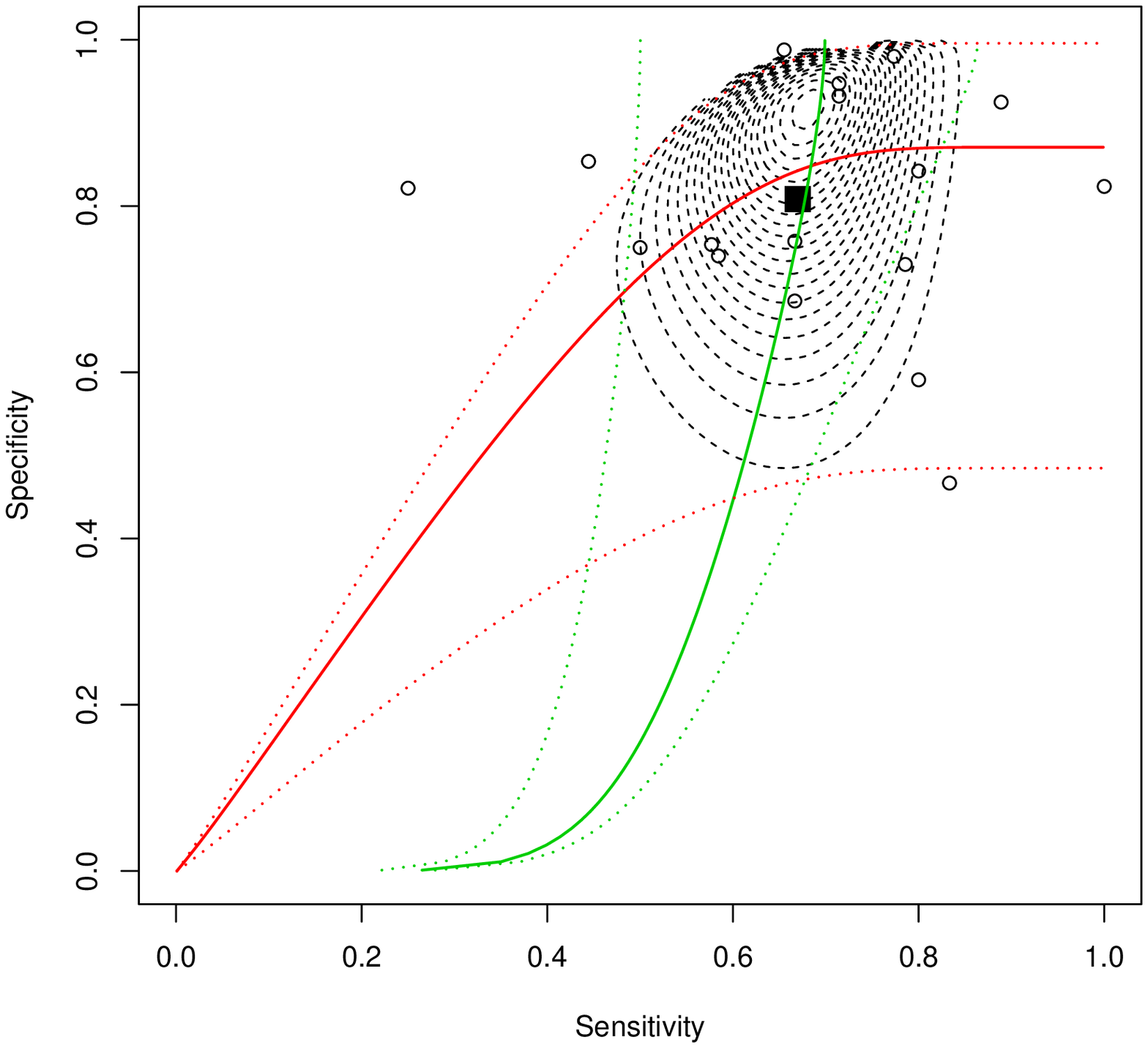}\\\hline
\end{tabular}
\end{footnotesize}
\caption{\label{SROC-LAG-beta}Contour plots and quantile  regression curves  from the copula representation of the random effects distribution with beta  margins and BVN, Frank, and Clayton by 180 and 270  copulas for the lymphangiography data. Red and green lines represent the quantile  regression curves $x_1:=\widetilde{x}_1(x_2,q)$ and $x_2:=\widetilde{x}_2(x_1,q)$, respectively; for $q=0.5$ solid lines and for $q\in\{0.01,0.99\}$ dotted lines.}
\end{center}
\vspace{-0.5cm}
\end{figure}

In Table \ref{lag-res} we report the resulting maximized log-likelihoods, estimates, and standard errors of the copula mixed models with different choices of parametric families of copulas and margins.
All models agree on the estimated sensitivity $\hat\pi_1$, but the estimate $\hat\pi_2$  of specificity is smaller when beta margins are assumed. The log-likelihoods show that a copula mixed model with rotated by 270 degrees Clayton copula and beta margins provides the best fit. It is revealed that a copula mixed model with the sensitivity and specificity on the original scale provides better fit than the GLMM, which models the sensitivity and specificity on a transformed scale.  The improvement over the GLMM is  small in terms of the likelihood principle, but for the Vuong's statistic there is enough improvement to get a statistical significant difference ($p$-value$= 0.025$).

Figures \ref{SROC-LAG-normal} and \ref{SROC-LAG-beta} show  the fitted SROC curves along with their confidence and prediction regions for  the copula mixed models with normal and beta margins, respectively. Note that the predictive regions cover a greater range of specificity rather than sensitivity.

\subsection{The magnetic resonance imaging data}
In this section we apply the copula mixed models  to data on 10 studies of magnetic resonance imaging for the diagnosis of lymph node metastasis in women with cervical cancer, the last imaging technique in the meta-analysis in \cite{Scheidler-etal-1997}. The  size in each study ranges from 20 to 272. Diagnosis of metastatic disease by lymphangiography relies on nodal enlargement.

In Table \ref{mri-res} we report the resulting maximized log-likelihoods, estimates, and standard errors of the copula mixed models with different choices of parametric families of copulas and margins.
All models roughly agree on the estimated sensitivity $\hat\pi_1$ and specificity  $\hat\pi_2$, but both are slightly smaller when beta margins are assumed. The log-likelihoods show that a rotated by 270 degrees Clayton  copula mixed model with  normal or beta margins provides the best fit.  Although, the rotated by 270 degrees Clayton  copula  mixed model  provides better fit than the GLMM, the difference, according to Vuong's test, is not statistical significant ($p$-value=0.156).

\begin{table}[!h]
  \centering
  \caption{\label{mri-res}Maximised  log-likelihoods, estimates and standard errors (SE), along with the Vuong's statistics and $p$-values for the magnetic resonance imaging data.}
  \begin{footnotesize}
    \begin{tabular}{c|cc|cc|cc|cc}
    \hline
    \multicolumn{9}{c}{Normal margins} \\
    \hline
          & \multicolumn{2}{c|}{BVN} & \multicolumn{2}{c|}{Frank} & \multicolumn{2}{c|}{Clayton by 90} & \multicolumn{2}{c}{Clayton by 270} \\
                      &   Estimate &         SE &   Estimate &         SE &   Estimate &         SE &   Estimate &         SE \\
\hline

    $\pi_1$ & 0.55  & 0.11  & 0.54  & 0.10  & 0.54  & 0.11  & 0.55  & 0.10 \\
    $\pi_2$ & 0.95  & 0.02  & 0.96  & 0.02  & 0.95  & 0.02  & 0.96  & 0.02 \\
    $\s_1$ & 1.16  & 0.39  & 1.14  & 0.38  & 1.21  & 0.41  & 1.13  & 0.37 \\
    $\s_2$ & 0.87  & 0.34  & 0.83  & 0.32  & 0.85  & 0.34  & 0.87  & 0.32 \\
    $\tau$ & -0.51 & 0.29  & -0.47 & 0.28  & -0.48 & 0.33  & -0.49 & 0.26 \\\hline
    $\log L$ & \multicolumn{2}{c}{-46.26} & \multicolumn{2}{c}{-46.35} & \multicolumn{2}{c}{-46.72} & \multicolumn{2}{c}{-45.90} \\\hline
    \multicolumn{9}{c}{Vuong's test} \\\hline
 $\sqrt{N}\bar D/s$        & \multicolumn{2}{c}{-} & \multicolumn{2}{c}{-0.815} & \multicolumn{2}{c}{-2.175} & \multicolumn{2}{c}{1.419} \\
 $p$-value          & \multicolumn{2}{c}{-} &\multicolumn{2}{c}{0.415} & \multicolumn{2}{c}{0.030} & \multicolumn{2}{c}{0.156} \\\hline
    \multicolumn{9}{c}{Beta margins} \\\hline
          & \multicolumn{2}{c|}{BVN} & \multicolumn{2}{c|}{Frank} & \multicolumn{2}{c|}{Clayton by 90} & \multicolumn{2}{c}{Clayton by 270} \\
                      &   Estimate &         SE &   Estimate &         SE &   Estimate &         SE &   Estimate &         SE \\
\hline
    $\pi_1$ & 0.54  & 0.08  & 0.53  & 0.08  & 0.53  & 0.08  & 0.54  & 0.08 \\
    $\pi_2$ & 0.94  & 0.02  & 0.94  & 0.02  & 0.94  & 0.02  & 0.94  & 0.02 \\
    $\g_1$ & 0.21  & 0.10  & 0.21  & 0.10  & 0.22  & 0.10  & 0.21  & 0.09 \\
    $\g_2$ & 0.04  & 0.03  & 0.03  & 0.02  & 0.03  & 0.03  & 0.04  & 0.02 \\
    $\tau$ & -0.53 & 0.28  & -0.47 & 0.28  & -0.50 & 0.33  & -0.50 & 0.25 \\\hline
   $\log L$ & \multicolumn{2}{c}{-46.27} & \multicolumn{2}{c}{-46.39} & \multicolumn{2}{c}{-46.75} & \multicolumn{2}{c}{-45.86} \\\hline
    \multicolumn{9}{c}{Vuong's test} \\\hline
     $\sqrt{N}\bar D/s$     & \multicolumn{2}{c}{-0.014} & \multicolumn{2}{c}{-0.422} & \multicolumn{2}{c}{-1.326} & \multicolumn{2}{c}{0.935} \\
   $p$-value        & \multicolumn{2}{c}{0.989} & \multicolumn{2}{c}{0.673} & \multicolumn{2}{c}{0.185} & \multicolumn{2}{c}{0.350} \\
    \hline
    \end{tabular}
 \end{footnotesize}
\end{table}

\begin{figure}[!h]
\begin{center}
\begin{footnotesize}
\begin{tabular}{|cc|}

\hline BVN& Frank\\\hline

\includegraphics[width=0.3\textwidth]{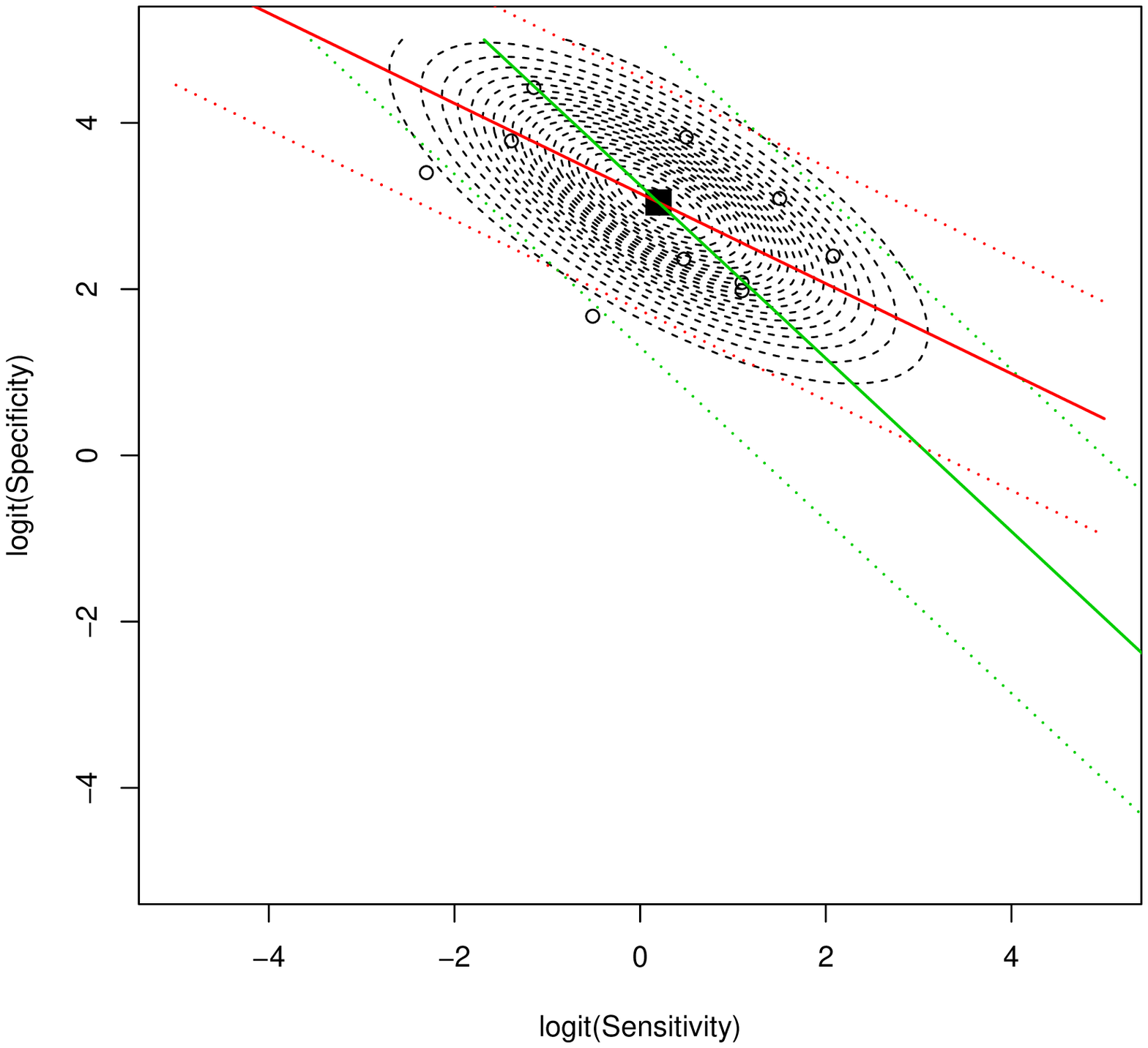}
&

\includegraphics[width=0.3\textwidth]{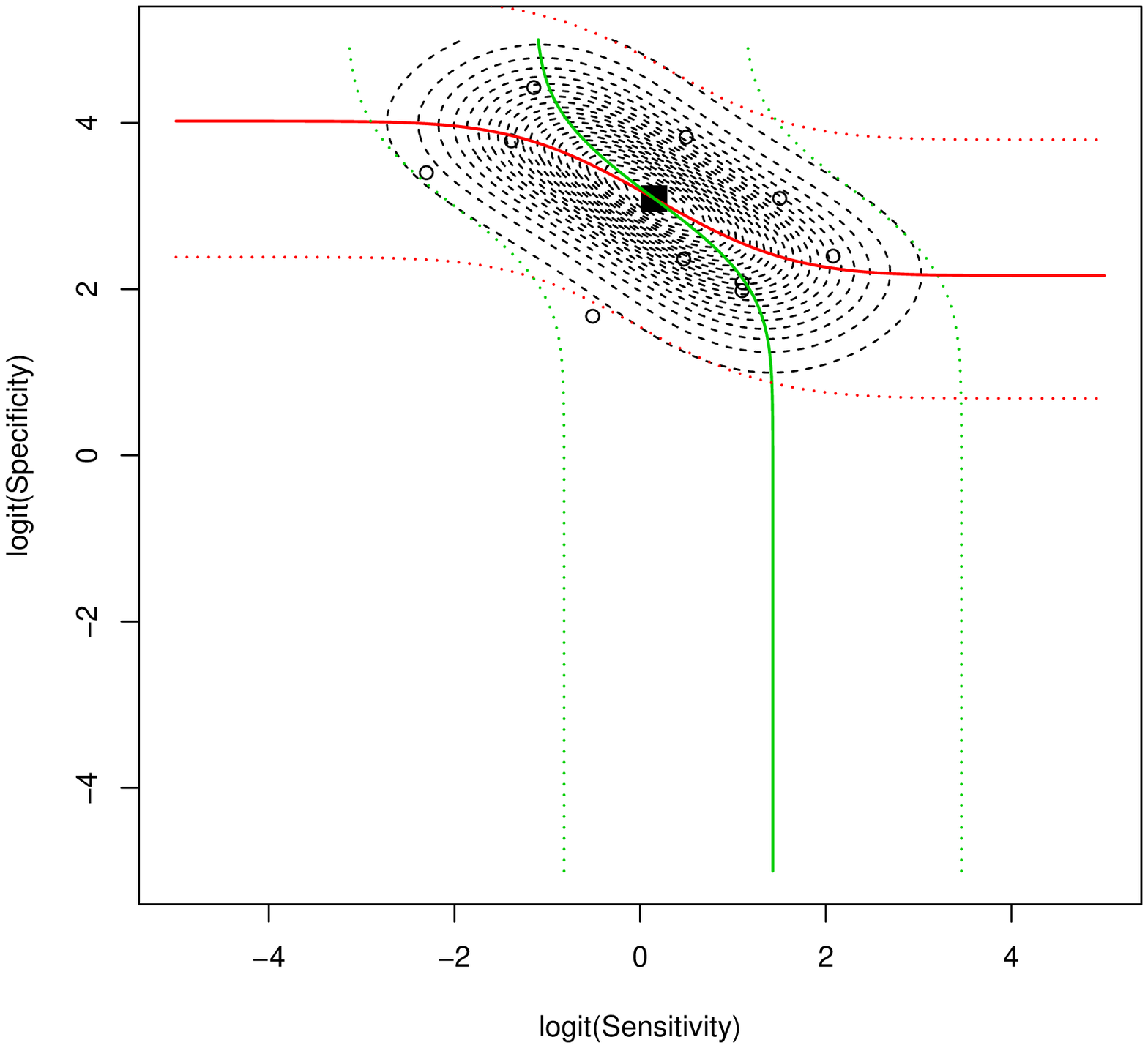}\\\hline
Clayton by 90& Clayton by 270\\\hline
\includegraphics[width=0.3\textwidth]{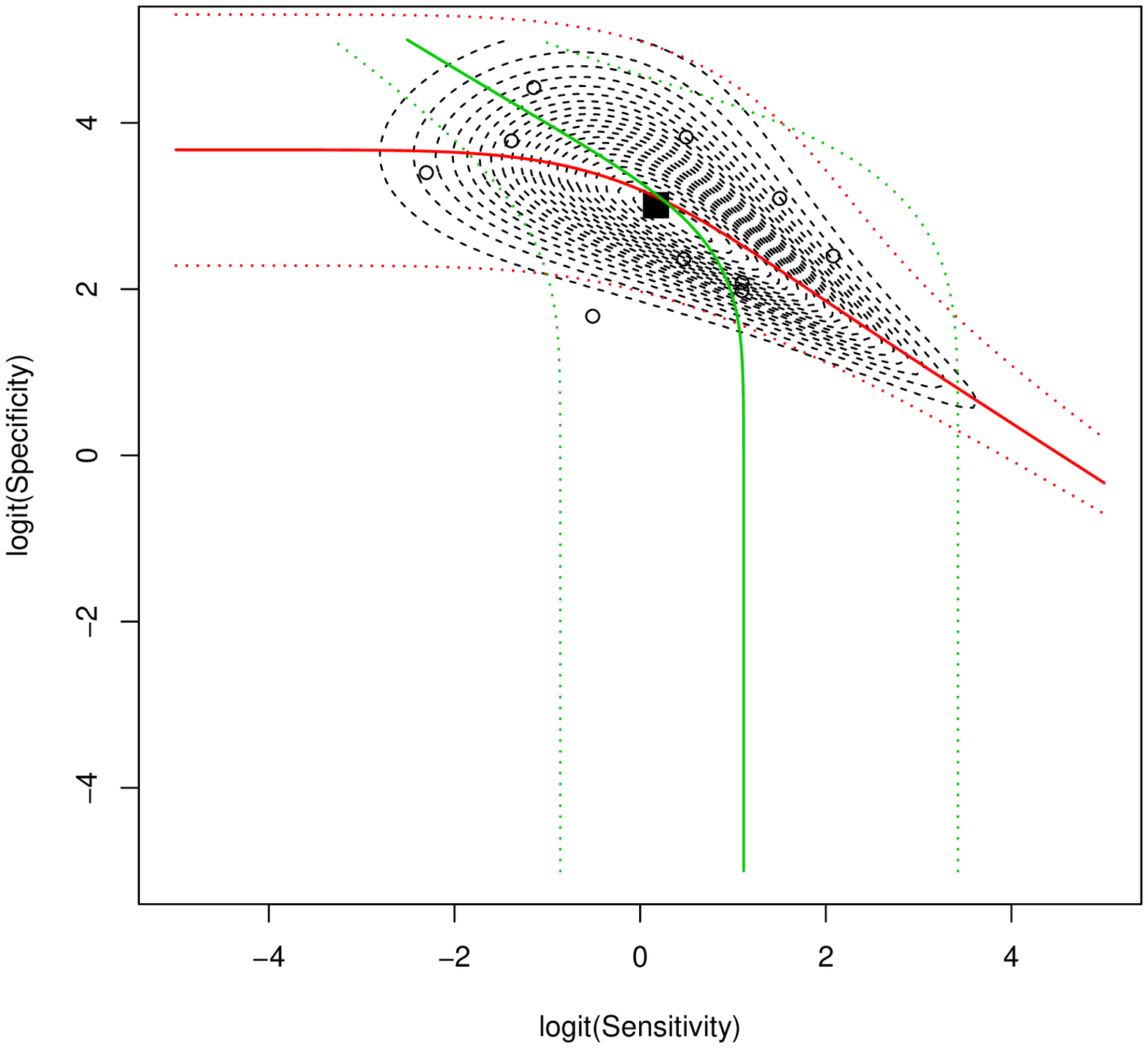}
&
\includegraphics[width=0.3\textwidth]{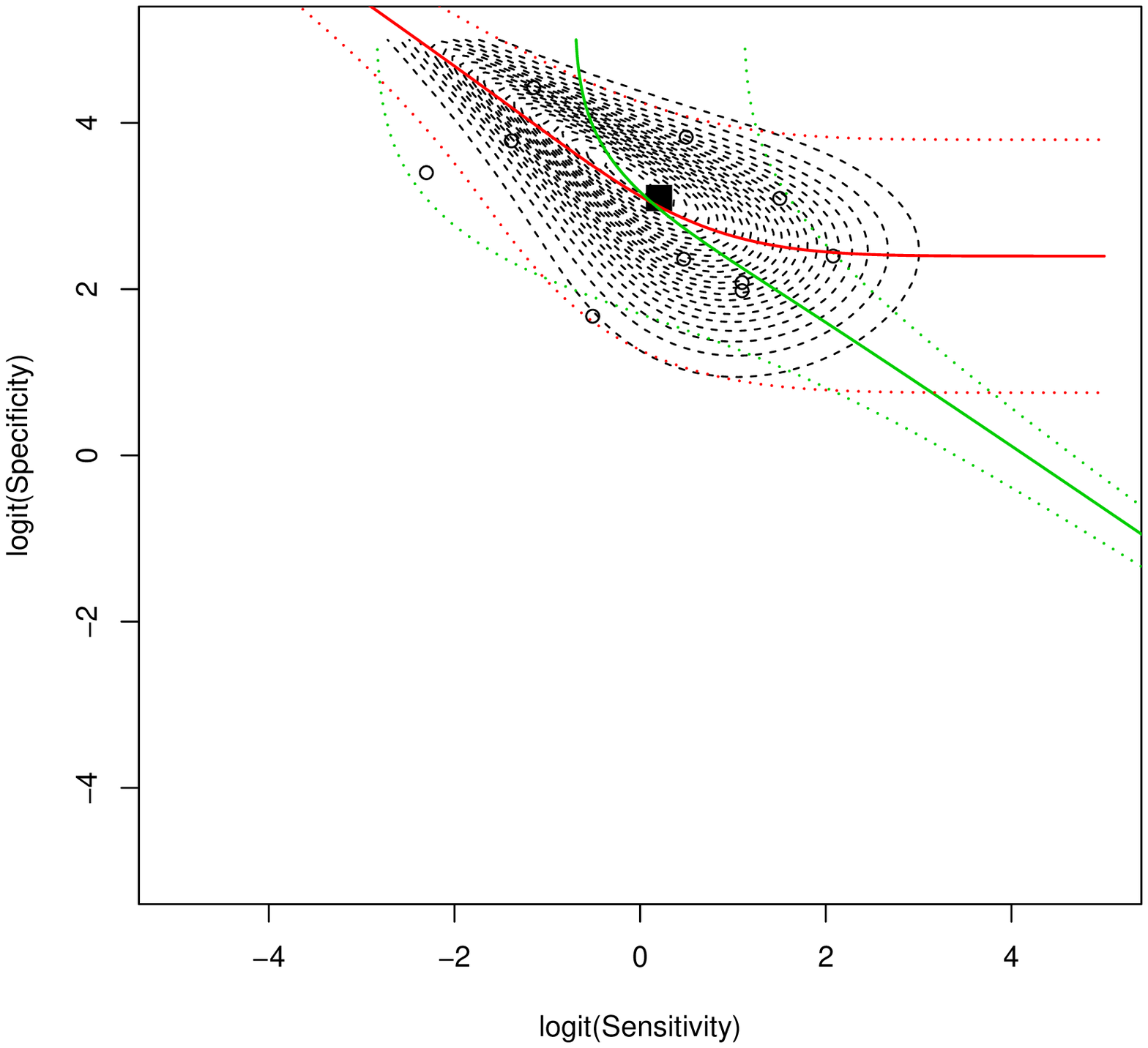}\\\hline
\end{tabular}
\end{footnotesize}
\caption{\label{SROC-normal-mri}Contour plots and quantile  regression curves  from the copula representation of the random effects distribution with normal  margins and BVN, Frank, and  Clayton by 90 and 270  copulas for the magnetic resonance imaging data. Red and green lines represent the quantile  regression curves $x_1:=\widetilde{x}_1(x_2,q)$ and $x_2:=\widetilde{x}_2(x_1,q)$, respectively; for $q=0.5$ solid lines and for $q\in\{0.01,0.99\}$ dotted lines.}
\end{center}
\vspace{-0.5cm}
\end{figure}

\begin{figure}[!h]
\begin{center}
\begin{footnotesize}
\begin{tabular}{|cc|}

\hline BVN& Frank\\\hline

\includegraphics[width=0.3\textwidth]{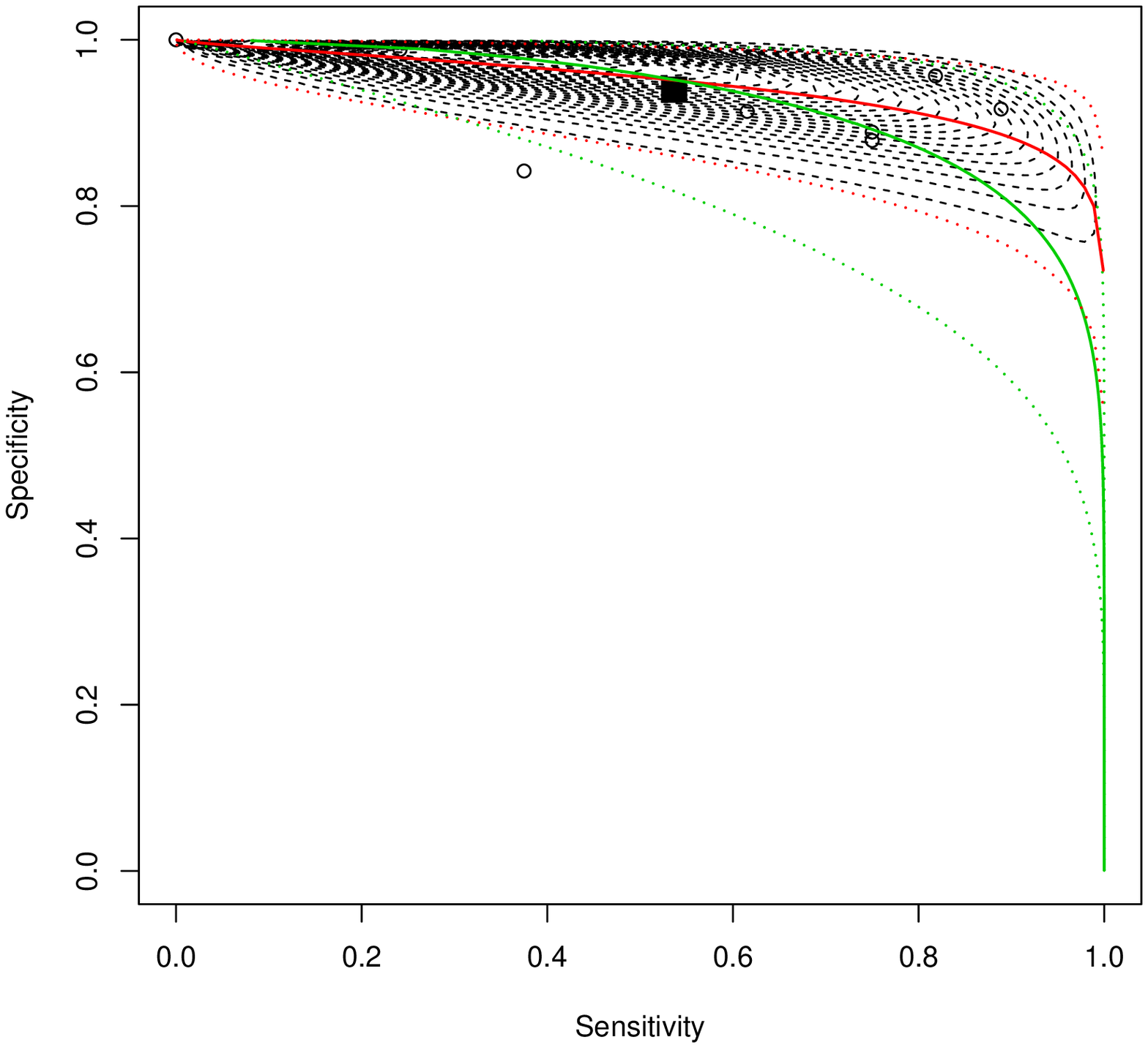}
&

\includegraphics[width=0.3\textwidth]{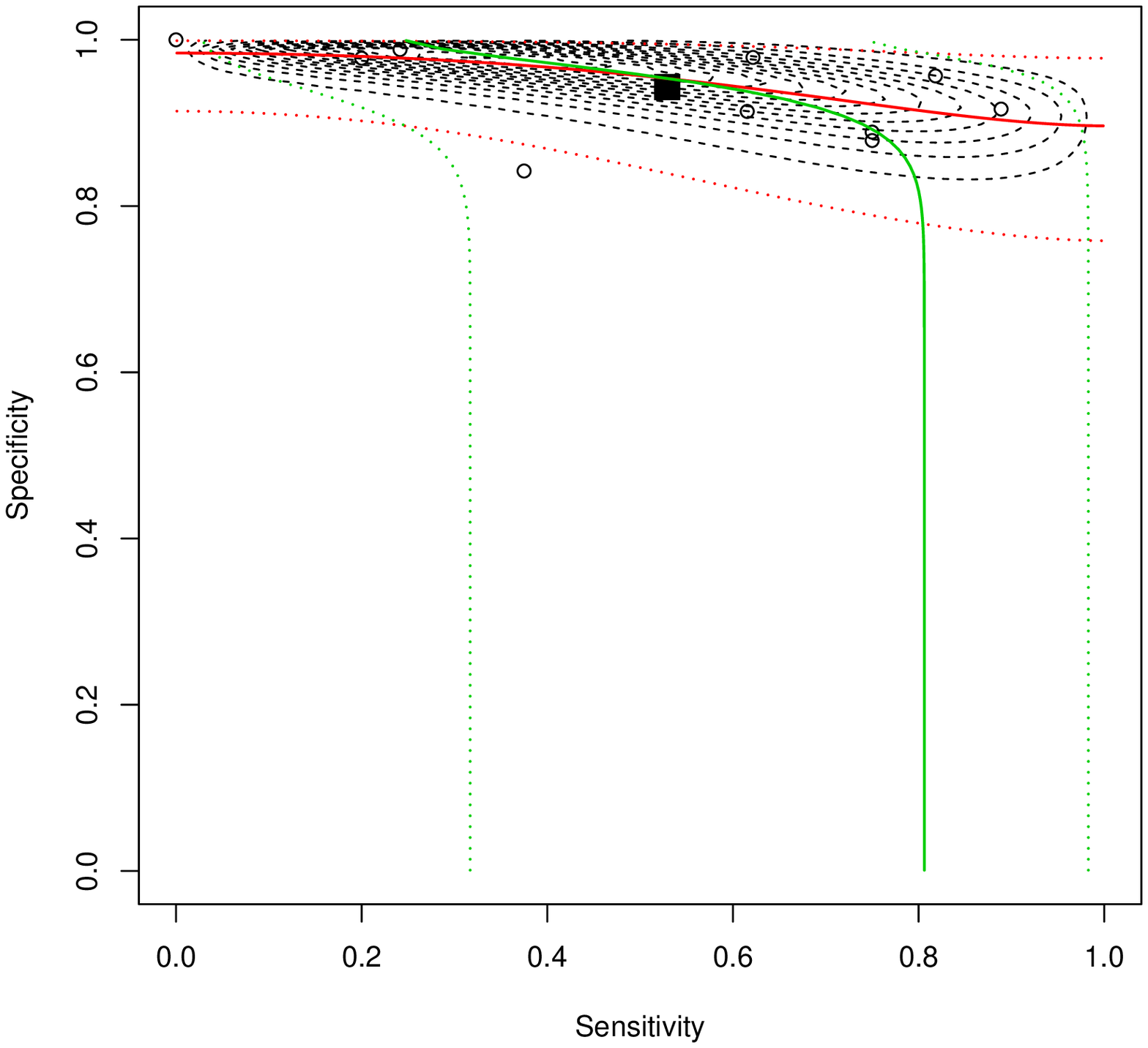}\\\hline
Clayton by 90& Clayton by 270\\\hline
\includegraphics[width=0.3\textwidth]{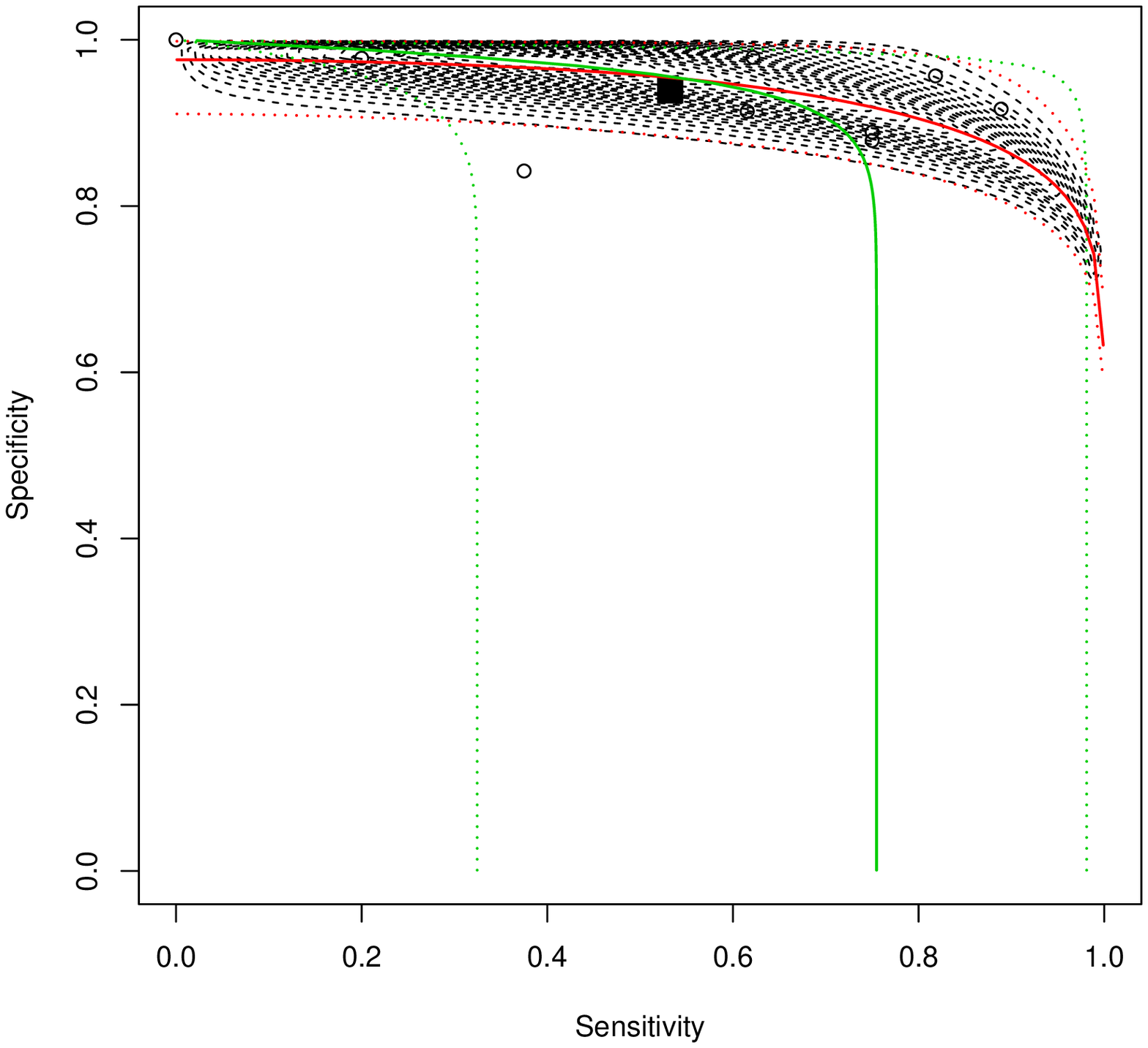}
&
\includegraphics[width=0.3\textwidth]{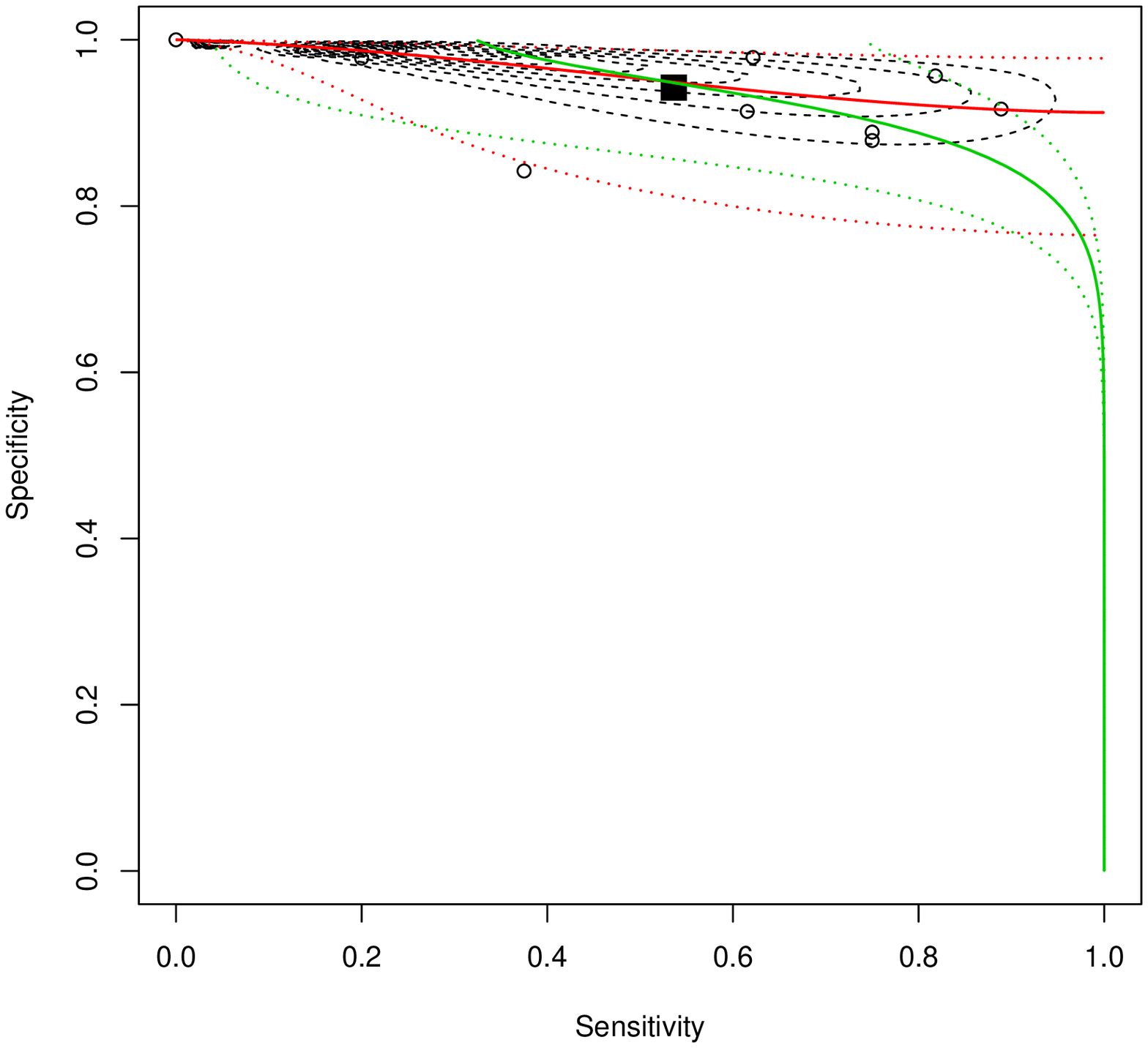}\\\hline
\end{tabular}
\end{footnotesize}
\caption{\label{SROC-beta-mri}Contour plots and quantile  regression curves  from the copula representation of the random effects distribution with beta  margins and BVN, Frank, and Clayton by 90 and 270  copulas for the magnetic resonance imaging data. Red and green lines represent the quantile  regression curves $x_1:=\widetilde{x}_1(x_2,q)$ and $x_2:=\widetilde{x}_2(x_1,q)$, respectively; for $q=0.5$ solid lines and for $q\in\{0.01,0.99\}$ dotted lines.}
\end{center}
\vspace{-0.5cm}
\end{figure}

Figures \ref{SROC-normal-mri} and \ref{SROC-beta-mri}  show  the fitted SROC curves along with their confidence and prediction regions for  the copula mixed models with normal and beta margins, respectively. Note that the predictive regions cover a greater range of sensitivity rather than specificity.


\section{\label{sec-discussion}Discussion}
We have proposed a copula mixed model for bivariate meta-analysis of diagnostic test accuracy studies.
This is the most general meta-analytic model, with univariate parameters separated from dependence parameters.
Our general model includes the GLMM as a special case  and  can provide an improvement over the latter based on log-likelihood and Vuong's \cite{vuong1989} statistic, and thus can provide a better statistical inference for the SROC.
This improvement relies on the
fact that the random effects distribution is expressed via copulas which allow for flexible dependence modelling, different from assuming simple linear correlation structures, normality and tail independence, which makes them well suited to the aforementioned application area.

Building on the basic model proposed in this paper, there are several extensions that can be implemented.
The copula mixed model   can also easily be extended in any context where clinical trials or observational studies report more than a single outcome and to inclusion of covariates.
However, larger sample sizes will be required to estimate the effect of covariates in bivariate meta-regression, where the underlying treatment effects depend on covariates. This is typical in the univariate meta-regression \cite{Jackson-2008}.

Another direction of future research is to extend our copula-based meta-analytic  model  to the $d$-variate ($d>2$) case.  There are many simple bivariate copula families, but generally
their multivariate extensions have limited dependence structures.
However, in recent years, a popular and useful approach is the vine pair-copula construction, see e.g.,
\cite{Kurowicka-Joe-2011,joe2014}, which is based on $d(d-1)/2$ bivariate copulas.
Some studies also may not report all $d$ outcomes. In such cases our model can be extended for missing data via pattern mixture models.
Pattern mixture models
are studied in \cite{shen-Weissfeld06} for copulas and in \cite{mavridis-etal-2014}   for pairwise and network meta-analysis.

\section*{\label{software}Software}

A contributed {\tt R} package {\tt CopulaREMADA} \cite{Nikoloulopoulos-2015}  has functions to implement the copula mixed model for meta-analysis of diagnostic test accuracy studies and produce SROC curves and summary operating points (a pair of average sensitivity and specificity) with a confidence region and a predictive region. All the analyses presented in Section \ref{sec-appl} are given as  code examples  in the package.
The {\tt R} package
{\tt VGAM} \cite{yee-2014} and specifically the functions {\tt pbetabinom} and {\tt dbetabinom} have been used  to implement the marginal distributions for the  KHS approximation method in \cite{kuss-etal-2013}.

\section*{Appendix}
We study the asymptotics of the KHS approximation method in \cite{kuss-etal-2013}, and we assess the accuracy based on the limit (as the number of clusters
increases to infinity) of the maximum KHS likelihood estimate  (KHSMLE).
By varying factors such as the marginal and copula parameters we demonstrate patterns
in the asymptotic bias of the KHSMLE, and assess the performance of KHS .
We will compute these limiting  KHSMLE  in a variety of situations to show clearly if  the KHS  method
is good. By using this limit, we show whether  or not this leads to consistent estimate of the parameters of
the bivariate random effects distribution; hence prove whether the KHS approach is valid or not.
For the cases where we compute the probability limit, we will take a constant size $n$ of groups of diseased and healthy people in the single studies that increases. For ease of exposition,  we also consider the case that the univariate marginal parameters are common to different univariate margins.

Let the $T$ distinct cases for the discrete response  be denoted as
$$(y_1^{(1)},y_2^{(1)}),\ldots, (y_1^{(T)},y_2^{(T)}).$$
In a random sample of size $N$, let the corresponding frequencies be denoted as $N^{(1)},\ldots, N^{(T)}$. Let $p^{(t)}$ be the limit in probability of $N^{(t)}/N$ as $N\to \infty$.

\setcounter{table}{0}

\renewcommand{\thetable}{A\arabic{table}}

\begin{table}[!h]
\caption{\label{kuss-asym}Limiting KHSMLE for a BVN copula mixed model with beta margins. }
\centering
\begin{footnotesize}
\begin{tabular}{ccc|ccc|ccc}
\hline
   $\rho$ & \multicolumn{ 2}{c}{$\rho^{KHS}$} &     $\pi$ & \multicolumn{ 2}{c}{$\pi^{KHS}$} &  $\gamma$ & \multicolumn{ 2}{c}{$\gamma^{KHS}$} \\\hline
           &    $n=20$ &   $n=100$ &            &    $n=20$ &   $n=100$ &            &    $n=20$ &   $n=100$ \\\hline

      -0.2 &     -0.047 &     -0.163 &        0.7 &      0.701 &      0.701 &       0.05 &      0.050 &      0.050 \\

      -0.5 &     -0.160 &     -0.412 &        0.7 &      0.705 &      0.702 &       0.05 &      0.049 &      0.050 \\

      -0.8 &     -0.277 &     -0.664 &        0.7 &      0.709 &      0.703 &       0.05 &      0.046 &      0.050 \\

     -1 &     -0.356 &     -0.825 &        0.7 &      0.711 &      0.704 &       0.05 &      0.044 &      0.049 \\

      -0.2 &     -0.051 &     -0.174 &        0.7 &      0.702 &      0.701 &        0.1 &      0.099 &      0.100 \\

      -0.5 &     -0.164 &     -0.438 &        0.7 &      0.708 &      0.703 &        0.1 &      0.095 &      0.099 \\

      -0.8 &     -0.289 &     -0.708 &        0.7 &      0.714 &      0.705 &        0.1 &      0.085 &      0.096 \\

     -1 &     -0.381 &     -0.885 &        0.7 &      0.718 &      0.706 &        0.1 &      0.075 &      0.089 \\

      -0.2 &     -0.030 &     -0.122 &        0.7 &      0.703 &      0.702 &        0.2 &      0.199 &      0.198 \\

      -0.5 &     -0.129 &     -0.320 &        0.7 &      0.715 &      0.707 &        0.2 &      0.187 &      0.187 \\

      -0.8 &     -0.242 &     -0.558 &        0.7 &      0.728 &      0.714 &        0.2 &      0.162 &      0.161 \\

     -1 &     -0.338 &     -0.788 &        0.7 &      0.738 &      0.722 &        0.2 &      0.133 &      0.116 \\

      -0.2 &     -0.007 &     -0.155 &        0.8 &      0.800 &      0.801 &       0.05 &      0.050 &      0.050 \\

      -0.5 &     -0.076 &     -0.396 &        0.8 &      0.804 &      0.802 &       0.05 &      0.049 &      0.049 \\

      -0.8 &     -0.148 &     -0.642 &        0.8 &      0.808 &      0.804 &       0.05 &      0.045 &      0.048 \\

     -1 &     -0.191 &     -0.802 &        0.8 &      0.811 &      0.805 &       0.05 &      0.041 &      0.045 \\

      -0.2 &     -0.005 &     -0.130 &        0.8 &      0.800 &      0.801 &        0.1 &      0.100 &      0.099 \\

      -0.5 &     -0.084 &     -0.339 &        0.8 &      0.808 &      0.804 &        0.1 &      0.095 &      0.094 \\

      -0.8 &     -0.168 &     -0.575 &        0.8 &      0.817 &      0.808 &        0.1 &      0.084 &      0.082 \\

     -1 &     -0.230 &     -0.765 &        0.8 &      0.823 &      0.813 &        0.1 &      0.073 &      0.064 \\

      -0.2 &      0.026 &     -0.062 &        0.8 &      0.795 &      0.803 &        0.2 &      0.202 &      0.196 \\

      -0.5 &     -0.064 &     -0.187 &        0.8 &      0.814 &      0.812 &        0.2 &      0.188 &      0.182 \\

      -0.8 &     -0.165 &     -0.348 &        0.8 &      0.837 &      0.825 &        0.2 &      0.156 &      0.148 \\

     -1 &     -0.248 &     -0.535 &        0.8 &      0.853 &      0.837 &        0.2 &      0.122 &      0.104 \\

      -0.2 &      0.084 &     -0.075 &        0.9 &      0.890 &      0.901 &       0.05 &      0.052 &      0.049 \\

      -0.5 &      0.031 &     -0.214 &        0.9 &      0.896 &      0.904 &       0.05 &      0.051 &      0.046 \\

      -0.8 &     -0.025 &     -0.377 &        0.9 &      0.903 &      0.907 &       0.05 &      0.048 &      0.039 \\

     -1 &     -0.062 &     -0.515 &        0.9 &      0.908 &      0.910 &       0.05 &      0.045 &      0.031 \\

      -0.2 &      0.114 &     -0.035 &        0.9 &      0.878 &      0.902 &        0.1 &      0.110 &      0.098 \\

      -0.5 &      0.045 &     -0.141 &        0.9 &      0.891 &      0.908 &        0.1 &      0.105 &      0.089 \\

      -0.8 &     -0.034 &     -0.269 &        0.9 &      0.908 &      0.916 &        0.1 &      0.092 &      0.071 \\

     -1 &     -0.092 &     -0.396 &        0.9 &      0.920 &      0.923 &        0.1 &      0.078 &      0.051 \\

      -0.2 &      0.182 &      0.032 &        0.9 &      0.837 &      0.893 &        0.2 &      0.241 &      0.193 \\

      -0.5 &      0.127 &     -0.054 &        0.9 &      0.858 &      0.911 &        0.2 &      0.225 &      0.168 \\

      -0.8 &      0.050 &     -0.179 &        0.9 &      0.894 &      0.936 &        0.2 &      0.178 &      0.115 \\

     -1 &     -0.032 &     -0.324 &        0.9 &      0.926 &      0.954 &        0.2 &      0.126 &      0.061 \\
\hline
\end{tabular}
\end{footnotesize}
\end{table}

For the KHS log-likelihood we have the limit,
\begin{equation}\label{limitACD}
N^{-1}\ell(\pi,\g,\th)\to \sum_{t=1}^{T} p^{(t)}
\log \Bigl[c\Bigl(H(y_{1}^{(t)};n,\pi,\g),H(y_{2}^{(t)};n,\pi,\g);\th\Bigr)\prod_{j=1}^2 h(y_{j}^{(t)};n,\pi,\g)\Bigr].
\end{equation}
The limit of the  KHSMLE (as $N\to \infty$) is the maximum of (\ref{limitACD});
we denote this limit as $(\pi^{KHS},\g^{KHS},\th^{KHS})$.
The $p^{(t)}$
in  (\ref{limitACD}) are the model based probabilities
and are computed to at least five significant digits using
Gauss-Legendre quadrature \cite{Stroud&Secrest1966} with a sufficient number of quadrature points as described in Subsection \ref{computation}.
For the log-likelihood in (\ref{beta-mixed-cop-likelihood}), we have the limit,
\begin{equation}\label{limit}
N^{-1}\ell(\pi,\g,\th)\to \sum_{t=1}^{T} p^{(t)}
\log \int\int
\prod_{j=1}^2g\Bigl(y_{j}^{(t)};n,F^{-1}(u_j;\pi,\g)\Bigr)c(u_1,u_2;\th)du_1du_2.
\end{equation}
The limit of the  MLE (as $N\to \infty$) is the maximum of (\ref{limit});
we denote this limit as $(\hat\pi, \hat\g, \hat\tau)$.

Representative results are shown in Table \ref{kuss-asym} for a BVN copula mixed model with beta margins,  with MLE results omitted because
they were identical with the true values up to four or five decimal places.
Therefore, our method leads to unbiased estimating
equations. Regarding the KHS method,
conclusions from the values in the table and other computations that we have done are that for the KHS method there is asymptotic bias (decreases as $n$  increases) for the univariate parameters $\pi$ and $\g$ as $\pi,\g$ and $\rho$ increase, and substantial asymptotic downward bias for the dependence parameter $\rho$; note that this slightly decreases as $n$  increases.

\section*{Acknowledgement}
Thanks to Professor Harry Joe, University of British Columbia, for insightful  comments.


\baselineskip=10pt

\begin{thebibliography}{10}
\itemsep=-5pt
\bibitem{Arends-etal-2008}
L.~R. Arends, T.~H. Hamza, J.~C. van Houwelingen, M.~H. Heijenbrok-Kal,
  M.~G.~M. Hunink, and T.~Stijnen.
\newblock Bivariate random effects meta-analysis of {R}{O}{C} curves.
\newblock {\em Medical Decision Making}, 28(5):621--638, 2008.

\bibitem{belgorodski10}
N.~Belgorodski.
\newblock {\em Selecting pair-copula families for regular vines with
  application to the multivariate analysis of European stock market indices}.
\newblock Diploma thesis, Technische Universitaet Muenchen, 2010.

\bibitem{Brechmann-Czado-Aas-2012}
E.~C. Brechmann, C.~Czado, and K.~Aas.
\newblock Truncated regular vines in high dimensions with applications to
  financial data.
\newblock {\em Canadian Journal of Statistics}, 40(1):68--85, 2012.

\bibitem{Chu&Cole2006}
H.~Chu and S.~R. Cole.
\newblock Bivariate meta-analysis of sensitivity and specificity with sparse
  data: a generalized linear mixed model approach.
\newblock {\em Journal of Clinical Epidemiology}, 59(12):1331--1332, 2006.

\bibitem{Chu&Guo2009}
H.~Chu and H.~Guo.
\newblock Letter to the editor.
\newblock {\em Biostatistics}, 10(1):201--203, 2009.

\bibitem{Chu-etal-2010}
H.~Chu, H.~Guo, and Y.~Zhou.
\newblock Bivariate random effects meta-analysis of diagnostic studies using
  generalized linear mixed models.
\newblock {\em Medical Decision Making}, 30(4):499--508, 2010.

\bibitem{Chu-etal-2012}
Haitao Chu, Lei Nie, Yong Chen, Yi~Huang, and Wei Sun.
\newblock Bivariate random effects models for meta-analysis of comparative
  studies with binary outcomes: Methods for the absolute risk difference and
  relative risk.
\newblock {\em Statistical Methods in Medical Research}, 21(6):621--633, 2012.

\bibitem{Demidenko04}
E.~Demidenko.
\newblock {\em Mixed {M}odels: {T}heory and {A}pplications}.
\newblock John Wiley \& Sons, Hoboken, New Jersey, 2004.

\bibitem{genest&johanna07}
C.~Genest and J.~Ne\v{s}lehov\'{a}.
\newblock A primer on copulas for count data.
\newblock {\em The {A}stin {B}ulletin}, 37:475--515, 2007.

\bibitem{genest&nikoloulopoulos&rivest08}
C.~Genest, A.~K. Nikoloulopoulos, L.-P. Rivest, and M.~Fortin.
\newblock Predicting dependent binary outcomes through logistic regressions and
  meta-elliptical copulas.
\newblock {\em Brazilian Journal of Probability and Statistics}, 27:265--284,
  2013.

\bibitem{genest87}
C.~Genest.
\newblock Frank's family of bivariate distributions.
\newblock {\em Biometrika}, 74(3):549--555, 1987.

\bibitem{genest&mackay86}
C.~Genest and J.~MacKay.
\newblock The joy of copulas: bivariate distributions with uniform marginals.
\newblock {\em The American Statistician}, 40(4):280--283, 1986.

\bibitem{glas-etal-2003}
A.S. Glas, D.~Roos, M.~Deutekom, A.H. Zwinderman, P.M. Bossuyt, and K.H. Kurth.
\newblock Tumor markers in the diagnosis of primary bladder cancer. a
  systematic review.
\newblock {\em The Journal of Urology}, 169(6):1975--1982, 2003.

\bibitem{hamza-etal-2009}
T.~H. Hamza, L.~R. Arends, H.~C. van Houwelingen, and T.~Stijnen.
\newblock Multivariate random effects meta-analysis of diagnostic tests with
  multiple thresholds.
\newblock {\em BMC Medical Research Methodology}, 9(1):1--15, 2009.

\bibitem{Harbord-etal-2007}
R.~M. Harbord, J.~J. Deeks, M.~Egger, P.~Whiting, and J.~A.~C. Sterne.
\newblock A unification of models for meta-analysis of diagnostic accuracy
  studies.
\newblock {\em Biostatistics}, 8(2):239--251, 2007.

\bibitem{Harbord&Whiting2009}
R.~M. Harbord and P.~Whiting.
\newblock metandi: {M}eta-analysis of diagnostic accuracy using hierarchical
  logistic regression.
\newblock {\em Stata Journal}, 9(2):211--229, 2009.

\bibitem{heffernan00}
J.~E. Heffernan.
\newblock A directory of coefficients of tail dependence.
\newblock {\em Extremes}, 3:279--290, 2000.

\bibitem{Hua-joe-11}
L.~Hua and H.~Joe.
\newblock Tail order and intermediate tail dependence of multivariate copulas.
\newblock {\em Journal of Multivariate Analysis}, 102(10):1454--1471, 2011.

\bibitem{HultLindskog02}
H.~Hult and F.~Lindskog.
\newblock {Multivariate extremes, aggregation and dependence in elliptical
  distributions}.
\newblock {\em {Advances in Applied Probability}}, {34}:{587--608}, {2002}.

\bibitem{Jackson-2008}
D.~Jackson.
\newblock The significance level of meta-regression's standard hypothesis test.
\newblock {\em Communications in Statistics - Theory and Methods},
  37(10):1576--1590, 2008.

\bibitem{JacksonRileyWhite2011}
D.~Jackson, R.~Riley, and I.~R.~White.
\newblock Multivariate meta-analysis: Potential and promise.
\newblock {\em Statistics in Medicine}, 30(20):2481--2498, 2011.

\bibitem{joe97}
H.~Joe.
\newblock {\em Multivariate {M}odels and {D}ependence {C}oncepts}.
\newblock Chapman \& Hall, London, 1997.

\bibitem{joe08}
H.~Joe.
\newblock Accuracy of laplace approximation for discrete response mixed models.
\newblock {\em Computational Statistics and Data Analysis}, 52(12):5066--5074,
  2008.

\bibitem{joe2010b}
H.~Joe.
\newblock Tail dependence in vine copulae.
\newblock In D~Kurowicka and H~Joe, editors, {\em Dependence Modeling: Vine
  Copula Handbook}, pages 165--187, Singapore, 2011. World Scientific.

\bibitem{joe2014}
H.~Joe.
\newblock {\em Dependence Modeling with Copulas}.
\newblock Chapman \& Hall, London, 2014.

\bibitem{Kurowicka-Joe-2011}
D.~Kurowicka and H.~Joe.
\newblock {\em Dependence {M}odeling -- {H}andbook on {V}ine {C}opulae}.
\newblock World Scientific Publishing Co, Singapore, 2011.

\bibitem{kuss-etal-2013}
O.~Kuss, A.~Hoyer, and A.~Solms.
\newblock Meta-analysis for diagnostic accuracy studies: a new statistical
  model using beta-binomial distributions and bivariate copulas.
\newblock {\em Statistics in Medicine}, 33(1):17--30, 2014.

\bibitem{Lee-1996}
T.~M.-L. Lee.
\newblock Properties and applications of the sarmanov family of bivariate
  distributions.
\newblock {\em Communications in Statistics - Theory and Methods},
  25(6):1207--1222, 1996.

\bibitem{Ma-etal-2013}
X.~Ma, L.~Nie, S.~R.~Cole, and H.~Chu.
\newblock Statistical methods for multivariate meta-analysis of diagnostic
  tests: An overview and tutorial.
\newblock {\em Statistical Methods in Medical Research}, 2013.

\bibitem{MavridisSalanti13}
D.~Mavridis and G.~Salanti.
\newblock A practical introduction to multivariate meta-analysis.
\newblock {\em Statistical Methods in Medical Research}, 22(2):133--158, 2013.

\bibitem{mavridis-etal-2014}
D.~Mavridis, I.~R.~White, J.~P.~T.~Higgins, A.~ Cipriani, and
  G.~Salanti.
\newblock Allowing for uncertainty due to missing continuous outcome data in
  pairwise and network meta-analysis.
\newblock {\em Statistics in Medicine}, 34(5):721--741, 2014.

\bibitem{nash90}
J.C. Nash.
\newblock {\em Compact Numerical Methods for Computers: Linear Algebra and
  Function Minimisation}.
\newblock Hilger, New York, 1990.
\newblock 2nd edition.

\bibitem{nelsen06}
R.~B. Nelsen.
\newblock {\em An {I}ntroduction to {C}opulas}.
\newblock Springer-Verlag, New York, 2006.

\bibitem{Nikoloulopoulos2013a}
A.~K. Nikoloulopoulos.
\newblock Copula-based models for multivariate discrete response data.
\newblock In F.~Durante, W.~H\"{a}rdle, and P.~Jaworski, editors, {\em Copulae
  in Mathematical and Quantitative Finance}, pages 231--249. Springer, 2013.

\bibitem{nikoloulopoulos13b}
A.~K. Nikoloulopoulos.
\newblock On the estimation of normal copula discrete regression models using
  the continuous extension and simulated likelihood.
\newblock {\em Journal of Statistical Planning and Inference}, 143:1923--1937,
  2013.

\bibitem{Nikoloulopoulos-2015}
A.~K. Nikoloulopoulos.
\newblock {\em {CopulaREMADA}: Copula random effects model for bivariate
  meta-analysis of diagnostic test accuracy studies}, 2015.
\newblock {R} package version 0.5.

\bibitem{nikoloulopoulos2015}
A.~K. {Nikoloulopoulos}.
\newblock {Efficient estimation of high-dimensional multivariate normal copula
  models with discrete spatial responses}.
\newblock {\em Stochastic Environmental Research and Risk Assessment}, 2015.
\newblock Accepted.

\bibitem{nikoloulopoulos&joe12}
A.~K. Nikoloulopoulos and H.~Joe.
\newblock Factor copula models for item response data.
\newblock {\em Psychometrika}, 2013.
\newblock
  \href{http://dx.doi.org/10.1007/s11336-013-9387-4}{http://dx.doi.org/10.1007/s11336-013-9387-4}.

\bibitem{Nikoloulopoulos&karlis08CSDA}
A.~K. Nikoloulopoulos and D.~Karlis.
\newblock Copula model evaluation based on parametric bootstrap.
\newblock {\em Computational Statistics \& Data Analysis}, 52:3342--3353, 2008.

\bibitem{Nikoloulopoulos&karlis07BIN}
A.~K. Nikoloulopoulos and D.~Karlis.
\newblock Multivariate logit copula model with an application to dental data.
\newblock {\em Statistics in Medicine}, 27:6393--6406, 2008.

\bibitem{normand99}
S.~L. Normand.
\newblock Meta-analysis: formulating, evaluating, combining, and reporting.
\newblock {\em Statistics in Medicine}, 18(3):321--59, 1999.

\bibitem{paul-etal-2010}
M.~Paul, A.~Riebler, L.~M. Bachmann, H.~Rue, and L.~Held.
\newblock Bayesian bivariate meta-analysis of diagnostic test studies using
  integrated nested laplace approximations.
\newblock {\em Statistics in Medicine}, 29(12):1325--1339, 2010.

\bibitem{Reitsma-etal-2005}
J.~B. Reitsma, A.~S. Glas, A.~W.S. Rutjes, R.~J.P.M. Scholten,
  P.~M. Bossuyt, and A.~H. Zwinderman.
\newblock Bivariate analysis of sensitivity and specificity produces
  informative summary measures in diagnostic reviews.
\newblock {\em Journal of Clinical Epidemiology}, 58(10):982--990, 2005.

\bibitem{Riley-etal-2007}
R.D. Riley, K.R. Abrams, A.J. Sutton, P.C. Lambert, and J.R. Thompson.
\newblock Bivariate random-effects meta-analysis and the estimation of
  between-study correlation.
\newblock {\em BMC Medical Research Methodology}, 7, 2007.

\bibitem{Rucker-schumacher-2009}
G.~R\:ucker and M.~Schumacher.
\newblock Letter to the editor.
\newblock {\em Biostatistics}, 10(4):806--807, 2009.

\bibitem{RutterGatsonis2001}
C.~M. Rutter and C.~A. Gatsonis.
\newblock A hierarchical regression approach to meta-analysis of diagnostic
  test accuracy evaluations.
\newblock {\em Statistics in Medicine}, 20(19):2865--2884, 2001.

\bibitem{Sarmanov1966}
O.~V. Sarmanov.
\newblock Generalized normal correaltion and two-dimensional {F}r\'echet
  classes.
\newblock {\em Doklady AN SSSR}, 168(1):596--599, 1966.

\bibitem{Scheidler-etal-1997}
J.~Scheidler, H.~Hricak, K.~K. Yu, L.~Subak, and M.R. Segal.
\newblock Radiological evaluation of lymph node metastases in patients with
  cervical cancer: A meta-analysis.
\newblock {\em JAMA}, 278(13):1096--1101, 1997.

\bibitem{serfling80}
R.~J. Serfling.
\newblock {\em Approximation Theorems of Mathematical Statistics}.
\newblock Wiley, New York, 1980.

\bibitem{shen-Weissfeld06}
Changyu Shen and Lisa Weissfeld.
\newblock A copula model for repeated measurements with non-ignorable
  non-monotone missing outcome.
\newblock {\em Statistics in Medicine}, 25(14):2427--2440, 2006.

\bibitem{sklar1959}
M.~Sklar.
\newblock Fonctions de r\'epartition \`a {$n$} dimensions et leurs marges.
\newblock {\em Publications de l'Institut de Statistique de l'Universit\'e de
  Paris}, 8:229--231, 1959.

\bibitem{Stroud&Secrest1966}
A.~H. Stroud and D.~Secrest.
\newblock {\em Gaussian Quadrature Formulas}.
\newblock Prentice-Hall, Englewood Cliffs, NJ, 1966.

\bibitem{vuong1989}
Q.~H. Vuong.
\newblock Likelihood ratio tests for model selection and non-nested hypotheses.
\newblock {\em Econometrica}, 57(2):pp. 307--333, 1989.

\bibitem{yee-2014}
T.~W. Yee.
\newblock {\em { VGAM}: Vector Generalized Linear and Additive Models}, 2014.
\newblock {R} package version 0.9-6.

\end{thebibliography}
\end{document}